\documentclass[12pt,preprint]{emulateapj}
\usepackage{natbib}
\usepackage{color}
\usepackage{hyperref} 
\hypersetup{colorlinks=true,citecolor=blue}

\def \bea {\begin{eqnarray}}
\def \ena {\end{eqnarray}}               
\def \bee {\begin{equation}}
\def \ene {\end{equation}}
\def    \simlt  {\lower.5ex\hbox{$\; \buildrel < \over \sim \;$}}
\def    \simgt  {\lower.5ex\hbox{$\; \buildrel > \over \sim \;$}}

\def	\ltsim	{\simlt}

\newcommand     \mum    {\,\mu{\rm m}}  

\def	\cm		{\,{\rm {cm}}}

\def	\B		{{\rm B}}
\def	\D		{{\rm D}}
\def	\erg		{\,{\rm {ergs}}}

\def    \exp 		{\,{\rm {exp}}}
\def	\g		{\,{\rm g}}

\def	\G		{\,{\rm G}}
\def	\GHz		{\,{\rm {GHz}}}

\def	\K		{\,{\rm K}}

\def	\AU		{\,{\rm {AU}}}

\def	\s		{\,{\rm s}}

\def    \yr  		{\,{\rm {yr}}}

\def	\H		{\rm H}

\def	\xhat		{\hat{\bf x}}
\def	\yhat		{\hat{\bf y}}
\def	\zhat		{\hat{\bf z}}

\def	\ehat		{\hat{\bf e}}

\def    \Bv     	{\bf  B}

\def    \kv     	{\bf  k}

\def	\ba		{{\bf a}}
\def	\be		{{\bf e}}

\def	\bB		{{\bf B}}

\def	\bJ		{{\bf J}}
\def	\bk		{\kv}

\def   	\bQ  		{{\bf Q}}

\def	\gas		{\rm {gas}}
\def	\th		{\rm {th}}

\def	\d		{\rm d}
\def	\rad		{\rm {rad}}

\def	\eff		{\rm {eff}}

\def    \coll        	{\rm {coll}}

\def	\ISRF		{\rm {ISRF}}

\def	\Bar		{\rm {Bar}}

\def	\sp		{\rm {sp}}

\font\mib=cmmib10

\def\bGamma{\hbox{\mib\char"00}}

\def\bmu{\hbox{\mib\char"16}}

\def\bphi{\hbox{\mib\char"1E}}

\def\bomega{\hbox{\mib\char"21}}

\begin{document}
\shorttitle{MRAT alignment}
\shortauthors{Hoang \& Lazarian}
\title{A unified model of grain alignment: Radiative Alignment of Interstellar Grains with magnetic inclusions}
\author{Thiem Hoang\altaffilmark{1}, and A. Lazarian\altaffilmark{2}}
\altaffiltext{1}
{Canadian Institute for Theoretical Astrophysics, University of Toronto, 60 St. George Street, Toronto, ON M5S 3H8, Canada}
\altaffiltext{2}{Department of Astronomy, University of Wisconsin-Madison}
  
\begin{abstract}
The radiative torque (RAT) alignment of interstellar grains with ordinary paramagnetic susceptibilities has been supported by earlier studies. The alignment of such grains depends on the so-called RAT parameter $q^{\max}$ that is determined by the grain shape. In this paper, we elaborate our model of RAT alignment for grains with enhanced magnetic susceptibility due to iron inclusions, such that RAT alignment is magnetically enhanced for which we term MRAT mechanism. Such grains can get aligned with high angular momentum at the so-called high-J attractor points, achieving a high degree of alignment. Using our analytical model  of RATs we derive the critical value of the magnetic relaxation parameter $\delta_{m}$ to produce high-J attractor points as functions of $q^{\max}$ and the anisotropic radiation angle relative to the magnetic field $\psi$. We find that if about $10\%$ of total iron abundance present in silicate grains are forming iron clusters, it is sufficient to produce high-J attractor points for all reasonable values of $q^{\max}$. To calculate the degree of grain alignment, we carry out numerical simulations of MRAT alignment by including stochastic excitations from gas collisions and magnetic fluctuations. We show that large grains can achieve perfect alignment when the high-J attractor point is present, regardless of the values of $q^{\max}$. Our obtained results pave the way for physical modeling of polarized thermal dust emission as well as magnetic dipole emission. {We also find that millimeter-sized grains in accretion disks may be aligned with the magnetic field if they are incorporated with iron nanoparticles.}
\end{abstract}

\section{Introduction}\label{sec:intro}
Immediately after the discovery of starlight polarization, more than 60 years ago by \cite{Hall:1949p5890} and \cite{Hiltner:1949p5851}, the polarization was attributed to differential extinction by nonspherical dust grains aligned with interstellar magnetic fields. This alignment of grains opened a new window into studying the magnetic fields, including the magnetic field topology and strength through starlight polarization (\citealt{1951ApJ...114..206D}; \citealt{1953ApJ...118..113C}) and polarized thermal dust emission (\citealt{Hildebrand:1988p2566}), in various astrophysical environments. Moreover, polarized thermal emission from aligned grains is a significant Galactic foreground source contaminating cosmic microwave background radiation (\citealt{2009AIPC.1141..222D}; \citealt{2014arXiv1409.5738P}). 

The problem of grain alignment has proven to be one of the longest standing problems in astrophysics. Over the last 60 years, a number of grain alignment mechanisms have been proposed and quantified (see \citealt{2007JQSRT.106..225L} and \citealt{LAH15} for reviews).  Some substantial extensions or modifications were suggested to the initial paradigm of grain alignment based on the \cite{1951ApJ...114..206D} paramagnetic relaxation theory.  However, an alternative alignment paradigm, based on radiative torques (RATs), has now become the favored mechanism to explain grain alignment. This mechanism was initially proposed by \cite{1976Ap&SS..43..291D}, but was mostly ignored at the time of its introduction due to the limited ability to generate quantitative theoretical predictions. \cite{1996ApJ...470..551D} (DW96) and \cite{1997ApJ...480..633D} (DW97) reinvigorated the study of the RAT mechanism by developing a numerical method based on discrete dipole approximation to compute RATs for several irregular grain shapes. Further advancement of the theory was done in \cite{2003ApJ...589..289W}. The strength of the torques obtained made it impossible to ignore them, but questions about basic properties (e.g., direction, dependence on grain size and shape) of the alignment for grains of different shapes as well as degree of alignment remained. 

The quantitative study of RAT alignment was initiated in a series of papers, starting with \cite{2007MNRAS.378..910L} (henceforth LH07) where an analytical model (AMO) of RAT alignment was introduced. The analytical model was the basis for further theoretical studies in (\citealt{Lazarian:2007p2442}; \citealt{Lazarian:2008fw}; \citealt{Hoang:2008gb} (HL08); \citealt{{2009ApJ...695.1457H},{2009ApJ...697.1316H}}, hereafter HL09ab). 

In the case of steady motion where only rotational damping due to gas collisions is considered, (i.e., rotational excitations is disregarded), we found that RATs tend to align ordinary paramagnetic grains at attractor points with low magnitude of angular momentum (so-called low-$J$ attractor points), and/or attractor points with high angular momentum (so-called high-$J$ attractor points).  LH07 considered ordinary paramagnetic susceptibilities that are being aligned by RATs and determined the parameter space for the alignment with high-J attractor points. Grains may get perfectly aligned at high-J attractor points (HL08) provided that the anisotropic radiation field and therefore RATs were sufficiently strong. The parameter space for such alignment was identified for the combination of the ratio of the two components of the radiative torques $q^{\max}$ which depends on the grain shape, as well as  the angle $\psi$ between the direction of the radiation anisotropy and the magnetic field. The grains outside this parameter space were found to be aligned at the low-J attractor point and were shown to be not perfectly aligned. For this generic feature, the degree of radiative alignment can be modeled by a parameter $f_{hi}$--the fraction of grains aligned at high-J attractor points, which allows quantitative predictions of the polarization by RAT alignment for various astrophysical conditions \citealt{2014MNRAS.438..680H}; \cite{2015MNRAS.448.1178H}. 

Observational evidence for RAT alignment is numerous and becomes increasingly available (\citealt{2007ApJ...665..369A}; \citealt{2008ApJ...674..304W}; \citealt{2010ApJ...720.1045A}; \citealt{2011PASJ...63L..43M}; \citealt{2011A&A...534A..19A}; \citealt{2016arXiv160405305R}). { Some fundamental properties of RAT alignment are observationally tested. For instance, the dependence of alignment on anisotropy direction of radiation have been tested and confirmed by observations (\citealt{2010ApJ...720.1045A}; \citealt{2011A&A...534A..19A}; \citealt{2015ApJ...812L...7V}). The loss of alignment toward the starless core because of the reduction of radiation intensity is observed by a number of groups (\citealt{2014A&A...569L...1A}; \citealt{Jones:2014fk})}. Evidence of enhancement of grain alignment by pinwheel torques was recently found by \cite{2013ApJ...775...84A} and modeled by \cite{2015MNRAS.448.1178H}. We refer interested readers to two recent reviews for extended discussions (\citealt{LAH15}; \citealt{Andersson:2015bq}).

Most of previous works on RAT alignment deal with ordinary paramagnetic grains (DW96; DW97; LH07; HL08; HL09ab), which is based on the assumption that iron atoms are distributed { diffusely} within silicate grains. However, iron atoms perhaps exist in the form of nanoparticles, and it is inevitably that some iron nanoparticles are incorporated in big grains. This can significantly enhance the grain magnetic susceptibility, and grains become superparamagnetic material \citep{Jones:1967p2924} (hereafter JS67). The first work that combines the effect of superparamagnetic inclusions and RATs was presented in \cite{Lazarian:2008fw} (hereafter LH08) where their joint action is found to convert a high-J repellor point to a high-J attractor point. This was not a trivial result of combining enhanced magnetic relaxation and suprathermal rotation induced by RATs.  Contrary to that, { in some realization, RATs would still induce grain alignment of high-J attractors without needing magnetic relaxation (see LH07)}. Nevertheless, we found that enhanced magnetic dissipation will tweak the RAT alignment as to stabilize the high-J attractor point, resulting in the increase in the degree of RAT alignment. In this paper, we will employ AMO to conduct an extensive study on {\it Magnetically enhanced RAT alignment}, which we term MRAT mechanism, aiming to identify the required enhancements of magnetic susceptibilities to produce high-J attractor points. 

Modern cosmology with Cosmic Microwave Background (CMB) polarization requires an accurate model of thermal dust polarization to enable reliable separation of galactic foreground contamination from the polarized CMB signal. We believe that an accurate model of polarized dust emission is achieved only when it is based on solid physics of grain alignment and tested physical properties of interstellar dust. Therefore, it is crucial to calculate the degree of grain alignment to be used for modeling of foreground polarization (cf. \citealt{Draine:2009p3780}). In addition, understanding of the grain alignment is necessary for reliably estimating magnetic fields with the Chandrasekhar - Fermi technique (see \citealt{2008ApJ...679..537F} for a newer edition of the technique).

{ Recent Planck results \citep{2015A&A...576A.104P} show a high degree of dust polarization, up to $20\%$, for the diffuse interstellar medium (ISM). Inverse modeling of starlight polarization usually requires perfect alignment of silicate grains to reproduce the maximum polarization degree (\citealt{Draine:2009p3780}; \citealt{2014ApJ...790....6H}, hereafter HLM14). This raises some challenge for the traditional RAT mechanism.
Indeed, the quantitative studies of radiative torques in LH07 and HL08 show that in the absence of high-J attractor points the degree of alignment is not expected to be close to 100$\%$. Our works show that the parameter space in terms of $q^{\max}$ angle $\phi$ for the high-J alignment is limited and therefore we expect a significant fraction of dust grains to be aligned with low-$J$ attractor point. The alignment in this case is unlikely to exceed $30\%$ (HL08). In this situation, one might have difficulty to explain some cases where of the high degree of polarization reported in the literature. Additional alignment enhancement via pinwheel torques potentially can solve the problem (HL09a). An alternative way of enhancing the degree of RAT alignment through iron inclusions was made in LH08. There it was suggested that combining enhanced magnetic dissipation and RATs can produce the high-J attractor for a larger value of the $q^{\max}-\phi$ parameter space, which for sufficiently large values of enhancement mean that all grains may be perfectly aligned. In what follows we provide the detailed study of MRAT mechanism and discuss its implications. }

We would like to stress that, in addition to thermal dust emission, magnetic dipole emission (MDE, \citealt{1999ApJ...512..740D}; \citealt{2013ApJ...765..159D}) is a potential foreground contamination to CMB B-mode signal in the frequency range below $300 $GHz. The MDE spectrum depends closely on the fraction of iron inclusions within the grain. The latter should also affect the alignment of composite grains. Therefore, to obtain a realistic polarization spectrum of MDE, it is first to compute the degree of alignment for composite grains. {We will quantify the degree of MRAT alignment of grains with magnetic inclusions for the different grain sizes and various levels of magnetic susceptibilities.} It is noted that \cite{2016ApJ...821...91H} (hereafter HL16) have computed the degree of alignment of composite grains and free-flying nanoparticles in the absence of RATs.

The present paper is organized as follows. { Section \ref{sec:dynamic} presents the main dynamical timescales for grain rotation.} In Section \ref{sec:grainmag}, we briefly describe magnetic properties of grains with superparamagnetic and ferromagnetic inclusions and show that the effect of magnetic relaxation should be taken into account in the RAT alignment paradigm. In Section \ref{sec:AMO}, we will use AMO to derive the critical conditions of magnetic inclusions required for high-J attractor points for arbitrary angles $\psi$ and $q^{\max}$ in the steady motion case. In Section \ref{sec:numdet}, we use numerical simulations to confirm the analytical results in the previous section. In Section \ref{sec:numstoc}, we perform numerical simulations to compute the degree of grain alignment for an extended parameter space of RATs, magnetic relaxation, and grain sizes. Discussion and summary are presented in Section \ref{sec:disc} and \ref{sec:sum}, respectively.

\section{Rotational Characteristic Timescales}\label{sec:dynamic}
\subsection{Rotational Damping by gaseous collisions}
For typical interstellar grains under interest here, the damping of grain rotation is dominated by collisions between the grain and gaseous atoms. For smaller grains, the damping by infrared emission and ion collisions can be important (see \citealt{1998ApJ...508..157D}; \citealt{Hoang:2010jy}; HL16).

The characteristic timescale of the rotational damping due to gas bombardment is given by
\bea
\tau_{\gas}\simeq 6.6\times 10^{4}\hat{\rho}\hat{s}^{-2/3}a_{-5}\left(\frac{300\K^{1/2}\cm^{-3}}{{T}_{\gas}^{1/2}n_{\H}}\right)\Gamma_{\|}^{-1} \yr,~~~\label{eq:tgas}
\ena
where $a_{-5}=a/10^{-5}$ cm with $a=a_{2}s^{1/3}$ the effective radius of the equivalent sphere of the same volume as the oblate spheroidal grain of semimajor axis $a_{2}$ and axial ratio $s<1$, $\hat{s}=s/0.5$, $\hat{\rho}=\rho/3\g\cm^{-3}$ with $\rho$ the mass density, $n_{\H}$ and $T_{\gas}$ are gas number density and temperature. { Here, $\Gamma_{\|}$ is the unity factor characterizing the grain geometry (see Appendix \ref{apdx:collexc})}.

\subsection{Magnetic Relaxation}
{ Iron atoms, if being distributed diffusely within a silicate grain, produce ordinary paramagnetic material. The zero-frequency susceptibility $\chi(0)$ of such a paramagnetic material is given by the Curie's law:
\bea
\chi(0)&=&\frac{n_{p}\mu^{2}}{3k_{\B}T_{\d}},\label{eq:curielaw}
\ena
where the effective magnetic moment per iron atom $\mu$ reads
\bea
\mu^{2}\equiv p^{2}\mu_{\B}^{2} =g_{e}^{2}\mu_{B}^{2}\left[{J(J+1)}\right],\label{eq:mu}
\ena
with $J$ being the angular momentum quantum number of electrons in the outer partially filled shell, and $p\approx 5.5$ is taken for silicate (see \citealt{Draine:1996p6977}).\footnote{For typical silicate of structure MgFeSi$O_{4}$, Fe$^{3+} (6S_{5/2})$ ion with $S=5/2, L=0$ and $J=5/2$ and $g_{e}\approx 2$ gives $p=g_{e} \sqrt{J(J+1)}=5.9$. For Fe$^{2+}(^{5}D_{4})$, one has $p=5.4$. So, we take $p\approx 5.5$ as a conservative value.} 

Plugging the typical numerical values into Equation (\ref{eq:curielaw}), we obtain
\bea
\chi(0)\simeq 0.03f_{p}\hat{n}_{23}\left(\frac{p}{5.5}\right)^{2}\left(\frac{20\K}{T_{d}}\right),\label{eq:chi_para}
\ena
where $\hat{n}_{23}=n/10^{23}\cm^{-3}$ is the atomic density of material, { $f_{p}$ is the fraction of paramagnetic (Fe) atoms in the dust grain}. For silicate of structure MgFeSiO$_{4}$ we have $f_{p}=1/7$ (see HLM14).}

A paramagnetic grain rotating with angular velocity $\bomega$ in an external magnetic field $\Bv$ experiences paramagnetic relaxation (\citealt{1951ApJ...114..206D}, henceforth DG51) that induces the dissipation of the grain rotational energy into heat. This results in the gradual alignment of $\bomega$ and angular momentum $\bJ$ with $\Bv$ at which the rotational energy is minimum. The classic DG51 model was originally suggested for the ordinary paramagnetic material, but it can be applied for any magnetic materials.

The characteristic time of the magnetic relaxation is given by 
\bea
\tau_{m} &=& \frac{I_{\|}}{K(\omega)VB^{2}}=\frac{2\rho a^{2}s^{-2/3}}{5K(\omega)B^{2}},\nonumber\\
&\simeq & 6\times 10^{5}\hat{\rho}\hat{s}^{-2/3}a_{-5}^{2}\hat{B}^{-2}\hat{K}^{-1} \yr,\label{eq:tau_DG_sup}
\ena
where $V=4\pi a^{3}/3$ is the grain volume, { $I_{\|}$ is the moment of inertia along the principal axis}, $\hat{B}=B/10\mu$G is the normalized magnetic field strength, and $\hat{K}=K(\omega)/10^{-13}\s$ and
$K(\omega)=\chi_{2}(\omega)/\omega$ with $\chi_{2}(\omega)$ is the imaginary part of complex magnetic susceptibility of the grain material.

To describe the aligning effect of magnetic relaxation relative to the disalignment by gas collisions, we introduce a dimensionless parameter
\bea
\delta_{m}&=&\frac{\tau_{\gas}}{\tau_{m}}\simeq 0.1\hat{B}^{2}\hat{K}{a}_{-5}\left(\frac{\hat{\rho}}{\hat{n}_{\H}\hat{T}_{\gas}^{1/2}}\right),\label{eq:deltam}
\ena
where $\hat{n}_{\H}=n_{\H}/(30\cm^{-3})$ and $\hat{T}_{\gas}=T_{\gas}/100\K$. { Above, the second-order effect of grain shape is ignored.}

The frequency-dependence $\chi_{2}(\omega)$ depends on the model. Following DL99, the critically-damped solution for $\chi_{2}(\omega)$ reads
\bea
\chi_{2}(\omega) = \frac{\chi(0)\tau_{2}\omega}{[1+(\omega \tau_{2}/2)^{2}]^{2}},\label{eq:chi2_cd}
\ena
{ where $\chi(0)$ is the magnetic susceptibility at $\omega=0$, and $\tau_{2}=2.9\times 10^{-12}/(f_{p}\hat{n}_{23})$ is the spin-spin relaxation time.}

From (\ref{eq:chi_para}) and (\ref{eq:chi2_cd}), we obtain
\bea
K(\omega) \simeq 8.7\times 10^{-14}\hat{p}\left(\frac{20\K}{T_{d}}\right)\frac{1}{[1+(\omega \tau_{2}/2)^{2}]^{2}},\label{eq:kappa_para}
\ena
where $\hat{p}=p/5.5$.


\subsection{Larmor precession}
A rotating paramagnetic grain can acquire magnetic moment through the Barnett effect (\citealt{Barnett:1915p6353}). The Rowland effect also produces some magnetic moment if the grain is electrically charged \citep{1971MNRAS.153..279M}. For paramagnetic material, the former is shown to be much stronger than that arising from the rotation of its charged body \citep{1976Ap&SS..43..291D}.

The instantaneous magnetic moment due to the Barnett effect is equal to
\bea
\bmu_{\Bar}=\frac{\chi(0)\bomega}{\gamma_{g}}V=-\frac{\chi(0)\hbar V}{g_{e}\mu_{B}}\bomega,\label{eq:muBar}
\ena
where $\gamma_{g}=-g_{e}\mu_{B}/\hbar\approx -e/(m_{e}c)$ is the gyromagnetic ratio of an electron, $g_{e}\approx 2$ is the $g-$factor, and $\mu_{\B}=e\hbar/2m_{e}c\approx 9.26\times 10^{-21} \erg \G^{-1}$ is the Bohr magneton. 

The interaction of the grain magnetic moment with an external static magnetic field $\bB$, governed by the torque $[\bmu_{\Bar}\times\bB]=-|\mu_{\Bar}|B\sin\beta\hat{\bphi}\equiv I_{\|}\omega\sin\beta d\phi/dt \hat{\bphi}$, causes the rapid precession of $\bJ\| \bomega$ around $\bB$. The period of such a Larmor precession denoted by $\tau_{\rm Lar}$, is given by
\bea
\tau_{\rm Lar}=\frac{2\pi}{d\phi/dt}=\frac{2\pi I_{\|}g\mu_{B}}{\chi_{0}V\hbar B}\simeq 0.65 a_{-5}^{2}\hat{s}^{-2/3}\frac{\hat{\rho}}{\hat{\chi}
\hat{B}} ~\yr,
\label{eq:tauB}
\ena
where $\hat{\chi}=\chi(0)/10^{-4}$. Comparing Equation (\ref{eq:tauB}) with Equations (\ref{eq:tgas}) and (\ref{eq:tau_DG_sup}), it is seen that the Larmor precession is much faster than the rotational damping and magnetic relaxation.

\section{Enhanced Magnetic Susceptibilities by Magnetic Inclusions}\label{sec:grainmag}
When iron atoms are incorporated in the form of clusters (e.g., nanoparticles), the magnetic susceptibility of the composite grain is substantially enhanced through superparamagnetic effect (JS67) and ferro-paramagnetic interactions \citep{1978ApJ...219L.129D}. Below, we briefly describe these effects for reference.

\subsection{Superparamagnetic Inclusions}
When ferromagnetic inclusions are sufficiently small so that thermal energy can exceed the energy barrier to reorient the inclusion magnetic moments, then thermal fluctuations within the grain can induce considerable fluctuations of the inclusion magnetic moments. { Let $N_{cl}$ be the number of iron atoms per inclusion (cluster).} In thermal equilibrium, the average magnetic moment of the ensemble of magnetic inclusions can be described by the Langevin function with argument $m H/kT_{d}$, where $m= N_{\rm cl}\mu_{0}$ { with $\mu_{0}$ being the magnetic moment per Fe atom in iron clusters} is the total magnetic moment of the cluster, and $H$ is the applied magnetic field (JS67). The composite grain exhibits superparamagnetic behavior, which has the magnetic susceptibility given by
\bea
\chi_{\sp}(0)=\frac{n_{\rm cl}m^{2}}{3kT_{d}},\label{eq:chisp}
\ena
where $n_{\rm cl}$ is the number of iron clusters per unit volume (JS67). 

Let $\phi_{\rm sp}$ be the volume filling factor of magnetic inclusions. Then, the total number of iron clusters per grain is
\bea
\mathcal{N}=3.5\times 10^{8}\phi_{\rm sp}N_{\rm cl}^{-1}a_{-5}^{3}. \label{eq:Ncl}
\ena

By plugging $m$ with $\mu_{0}=p\mu_{B}$ and $n_{\rm cl}=\mathcal{N}/V$ into Equation (\ref{eq:chisp}), we obtain:
\bea
\chi_{\sp}(0)\approx 0.026N_{\rm cl}\phi_{\rm sp}\hat{p}^{2}\hat{T}_{d}^{-1},\label{eq:super_chi}
\ena
where  $\hat{T}_{d}=T_{d}/20\K$.

Superparamagnetic inclusions undergo thermally activated remagnetization at rate
\bea
\tau^{-1}_{\sp}\approx \nu_0 \exp\left(-\frac{N_{\rm cl}T_{\rm act}}{T_{d}}\right)
\label{sp}
\ena
where { experiments give} $\nu_0\approx 10^{9}\s^{-1}$ and $T_{\rm act}\approx 0.011\K$ (see \citealt{Morrish:2001vp}). 

{ The frequency dependence susceptibility $\chi_{\rm sp}(\omega)$ is then evaluated using Equation (\ref{eq:chi2_cd}) with $\chi(0)=\chi_{\rm sp}(0)$ and $\tau_{2}=\tau_{\rm sp}$. Thus,}
\bea
K_{\sp}(\omega) &=& \frac{\chi_{\sp}(0)\tau_{\sp}}{[1+(\omega \tau_{\sp}/2)^{2}]^{2}},\nonumber\\
&\simeq & 2.6\times 10^{-11}N_{\rm cl}\phi_{\rm sp}\hat{p}^{2}\frac{\exp\left(N_{\rm cl}T_{\rm act}/T_{d}\right)}{\hat{T}_{d}[1+(\omega \tau_{\sp}/2)^{2}]^{2}}.~~~\label{eq:kappa_sp}
\ena

{ Here we have disregarded the magnetic susceptibility term by individual iron inclusions which are important only for high rotation frequency $\omega>10^{9}$ rad/s (see HL16).}

If the entire iron budget (also Mg, Si) is incorporated into dust to form a structure MgSiFeO$_{4}$, then the volume filling factor is $\phi_{\sp}=V_{\rm Fe}^{3}/(V_{\rm Mg}^{3}+V_{\rm Si}^{3}+V_{\rm Fe}^{3}+4V_{\rm O}^{3})=0.29$, { where $V_{j}$ is the volume of $j$ atom.} The typical value from observations of glass with embedded metal and sulfides (GEMS) is $\phi_{\sp}\approx 0.03$ (\citealt{Bradley:1994p6379}; \citealt{1995ApJ...445L..63M}), which indicates about $0.03/0.29\approx 10\%$ of Fe depleted into dust silicate.

In Figure \ref{fig:delta_m} we show the contours of $\delta_{m}$ in the plane of $N_{\rm cl}$ and $\phi_{\rm sp}$ for the $a=0.1\mum$ grains with magnetic inclusions, where $\phi_{\rm sp}$ ranges from $10^{-3}$ to $0.3$ (its maximum value) at which dust accommodates all Fe abundance. It can be seen that $\delta_{m}>1$ even for a small values of $N_{\rm cl}$ and $\phi_{\sp}$. Therefore, even with a small level of magnetic inclusions, the effect of magnetic relaxation on RAT alignment cannot be ignored. 

\subsection{Effect of ferromagnetic inclusions on paramagnetic atoms}
JS67 show that spin-lattice relaxation vanishes when there is no static magnetic field, which is true for rotating grains. Therefore, the spin-spin relaxation is the way to produce magnetic susceptibility and paramagnetic alignment (DG51). \cite{1978ApJ...219L.129D} pointed out that the existence of ferromagnetic inclusions in a big silicate grain can generate internal static magnetic fields that act on nearby paramagnetic (iron) atoms. As a result, the spin-lattice relaxation for paramagnetic atoms would exist and increase the magnetic susceptibility of the composite system. The total susceptibility is equal to
\bea
K(\omega) = \mathcal{F}\frac{\chi(0)\tau_{1}}{1+(\omega \tau_{1})^{2}} + (1- \mathcal{F})\frac{\chi(0)\tau_{2}}{1+(\omega \tau_{2})^{2}},\label{eq:K_ferropara}
\ena
where $\tau_{1}$ is the spin-lattice relaxation time. The first term is only present in the presence of the static internal field generated by iron clusters, and the factor $\mathcal{F}$ reads 
\bea
 \mathcal{F}=\frac{H_{0}^{2}}{H_{0}^{2}+H_{i}^{2}},
\ena
where $H_{0}$ is the rms value of the magnetic field due to iron clusters in a given direction, and $H_{i}$ is the internal field produced by paramagnetic atoms. The value of $H_{i}$ reads
\bea
 H_{i}=\left(\frac{C_{M}T_{d}}{\chi(0)} \right)^{1/2},
\ena
where $C_{M}$ is the heat volume capacity at constant magnetization. Various paramagnetic materials have $H_{i}\sim 10^{3}$ Oe (see \citealt{Draine:1996p6977}).

Let $f_{\rm Fe,SD}$ be the fraction of the total iron atoms present in single-domain ferromagnetic inclusions. Then, 
\bea
H_{0}&\simeq &3.8f_{\rm Fe,SD}np\mu_{B}/\sqrt{3},\nonumber\\
&\simeq& 110\left(\frac{f_{\rm Fe, SD}}{0.01}\right)\left(\frac{p}{5.5}\right)\hat{n}_{23}~ {\rm Oe}.
\ena

With $\tau_{1}\sim 10^{-6}\s$, the ferromagnetic-paramagnetic interaction can raise $K(\omega)$ by two orders of magnitude above the susceptibility of the normal paramagnetic material.

\begin{figure*}
\centering
\includegraphics[width=0.45\textwidth]{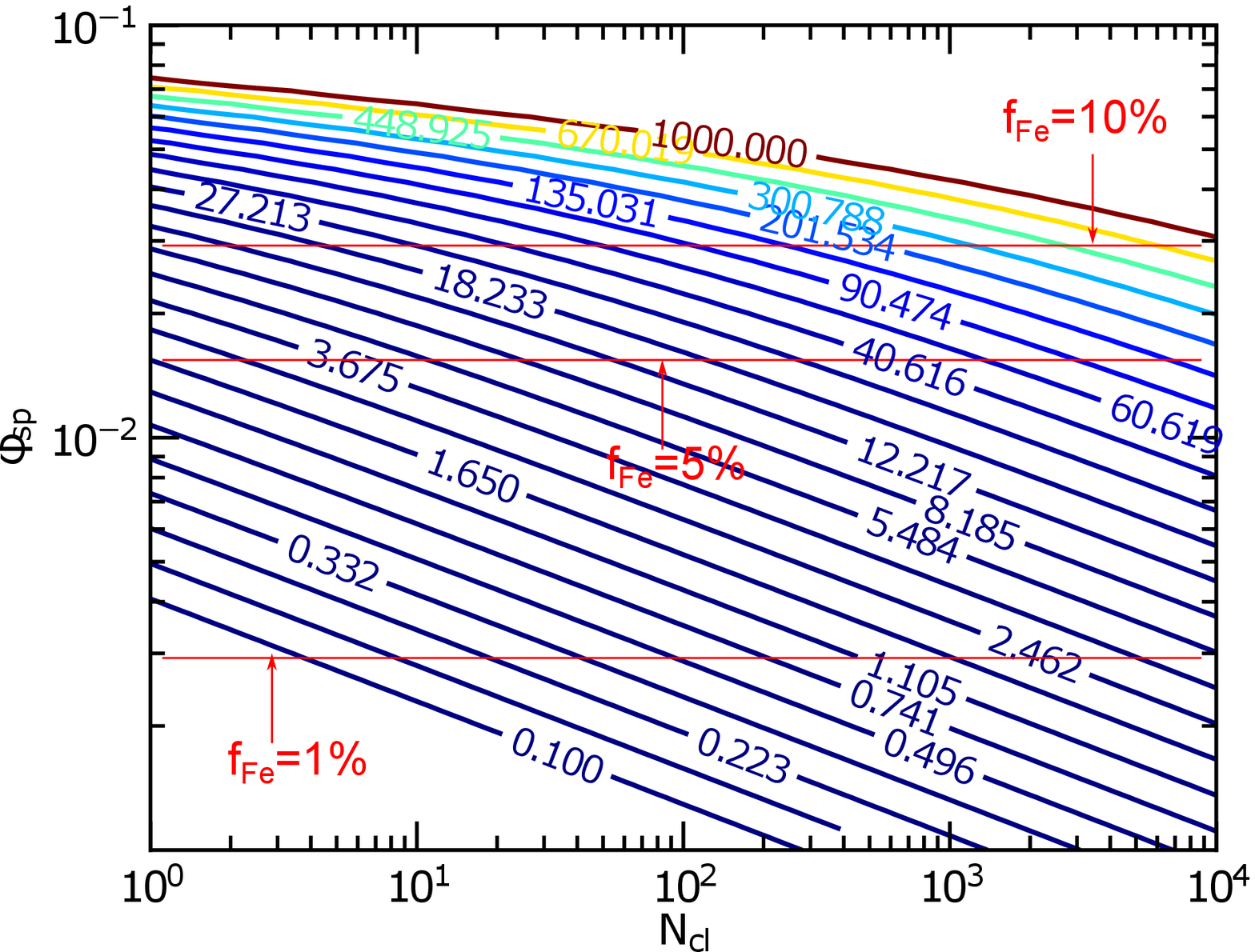}
\includegraphics[width=0.45\textwidth]{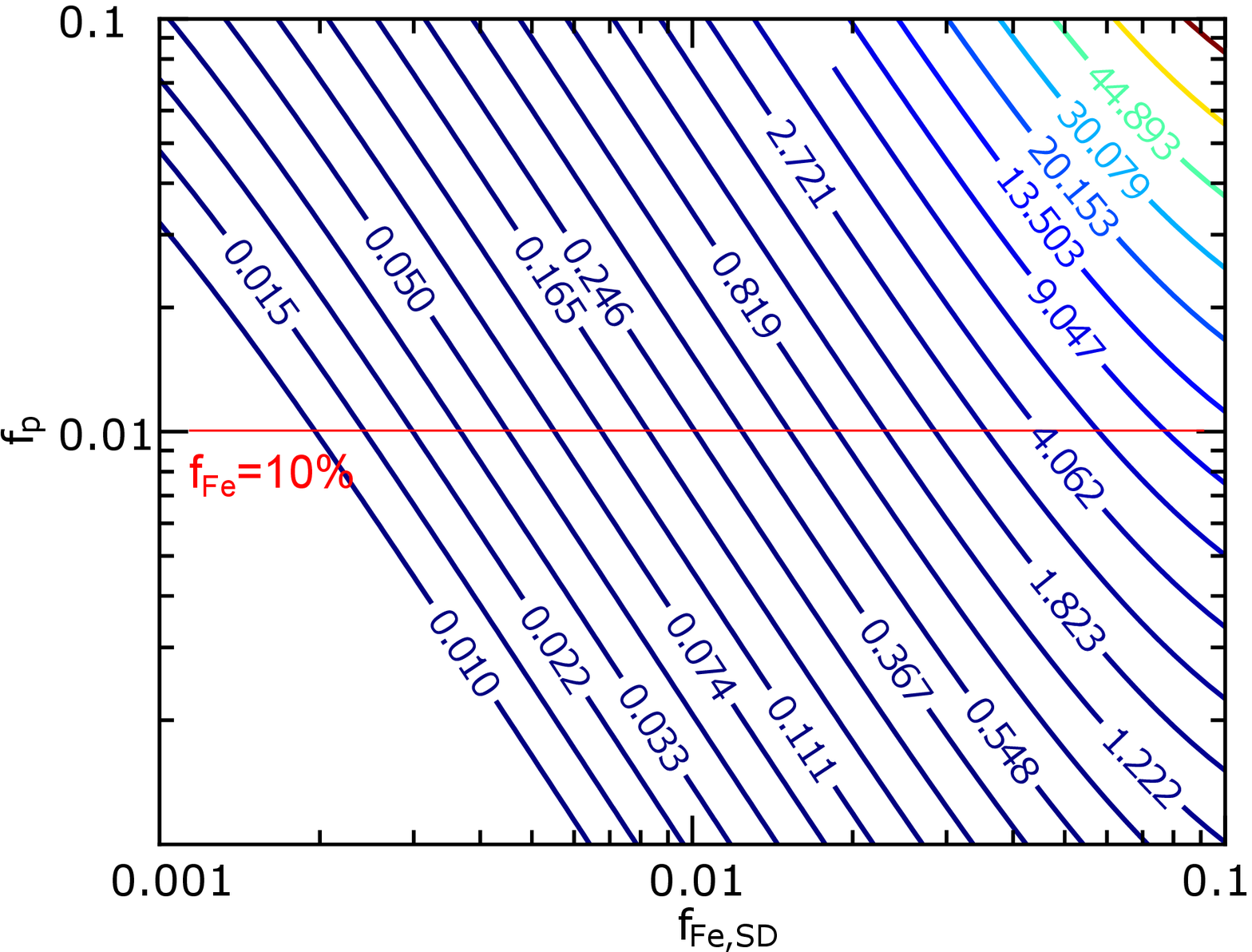}
\caption{Contour of the magnetic relaxation parameter, $\delta_{m}$, for superparamagnetic grains (left) and ferromagnetic-paramagnetic interaction (right). The grain size $a=0.1\mum$ and the physical parameters of the CNM are listed in Table \ref{tab:notation}. Horizontal lines mark the corresponding fraction of iron depleted into the dust, $f_{\rm Fe}$.}
\label{fig:delta_m}
\end{figure*}

Figure \ref{fig:delta_m} (right panel) shows contours of $\delta_{m}$ in the plane of $f_{\rm Fe,SD}$ and $f_{p}$ where $f_{p}$ ranges from $10^{-3}$ to $0.1$, corresponding to about $1\%$ to $100\%$ of Fe abundance depleted into dust. { It is noted that upto $80\%$ of Fe can be in dust (\citealt{2009ApJ...700.1299J}).} As expected, the ferro-para interaction can be important for larger values of $f_{\rm Fe,SD}$, but it is much weaker than the superparamagnetic effect (left panel). { Thus, in the following, we consider the superparamagnetic effect of iron inclusions only.}

\begin{table}[htb]
\begin{center}
\caption{\label{tab:notation}
Physical parameters and chose values for calculations}
\begin{tabular}{l l l}
\hline\hline
Symbol & Meaning  & Chosen values\cr
\hline
$n_{\H}$ & proton number density of gas & $30\cm^{-3}$ \cr 
$T_{\rm gas}$& gas temperature & 100 K\cr
$T_{\rm d}$ & dust temperature & 20 K \cr
$B$ & magnetic field strength & 10 $\mu$G\cr
$\rho$ & mass density of dust & $3\g\cm^{-3}$\cr
$\gamma$ & anisotropy of radiation field & 0.1\cr
$\bar{\lambda}$ & mean wavelength & $1.2\mum$\cr
$u_{\ISRF}$ & energy density of radiation & $8.64\times 10^{-13}\erg\cm^{-3}$\cr
\cr
\hline\hline\\
\end{tabular}
\end{center}
\end{table}

{ Table \ref{tab:notation} lists some important parameters and their meaning used in the paper.}

\section{Overview of AMO and Parameters of MRAT alignment}\label{sec:AMO}
In the following, we first review the fundamental features of our analytical model (AMO) of RATs.

\subsection{Overview of RATs and AMO}
Let $u_{\lambda}$ be the spectral energy density of radiation field at wavelength $\lambda$ and $\gamma$ be its anisotropy. The energy density of the radiation field is $u_{\rad}=\int u_{\lambda}d\lambda$. Radiative torque arising from the interaction of an anisotropic radiation field of direction $\bk$ with an irregular grain of size $a$ is then given by
\bea
{\bGamma}_{\lambda}=\gamma \pi a^{2}
u_{\lambda} \left(\frac{\lambda}{2\pi}\right){\bQ}_{\Gamma},\label{eq:GammaRAT}
\ena
where ${\bf Q}_{\Gamma}$ is the RAT efficiency that can be decomposed into three components $Q_{e1}, Q_{e2}$ and $Q_{e3}$ in the scattering reference frame defined by unit vectors $\ehat_{1},\ehat_{2},\ehat_{3}$ with $\hat{\be}_{1} \| \bk$, $\ehat_{2}\perp \ehat_{1}$ and $\ehat_{3}=\ehat_{1}\times \ehat_{2}$ (see the left panel of Figure \ref{fig:RFs}; see also DW96; LH07).

The magnitude of RAT efficiency, $Q_{\Gamma}$, in general depends on the radiation field, grain shape, size and grain orientation relative to $\bk$. When the anisotropic radiation direction is parallel to the axis of maximum moment of inertia $\hat{\ba}_{1}$, LH07 found that the magnitude of RAT efficiency by a monochromatic radiation of wavelength $\lambda$ for a grain of size $a$ can be approximated by a power-law as:
\bea
Q_{\Gamma}\sim 0.4\left(\frac{{\lambda}}{a}\right)^{\eta},\label{eq:QAMO}
\ena
where $\eta=0$ for $\lambda \ltsim 2a$  and $\eta=-3$ for $\lambda \gg a$. 

{ The exact RAT efficiency also depends on the dielectric function of dust (LH07). The presence of iron inclusions can modify the dielectric function of the composite dust, but it occurs mostly in the microwave frequency range (see DL99; DH13). Since RATs are determined by the interaction of optical-UV photons with the dust grain, the effect of iron inclusions on RATs is expected to be negligible.}

For the full spectrum of the diffuse interstellar radiation field (\citealt{1983A&A...128..212M}), the averaged torque magnitude is approximately given by
\bea
\bar{Q}_{\Gamma}\approx 2.4\times 10^{-3}\left(\frac{\bar{\lambda}}{1.2\mum}\right)^{-2.7}a_{-5}^{2.7},\label{eq:QRAT_avg}
\ena
for $a\ll \bar{\lambda}$, where $\bar{\lambda}=1.2\mum$ is the mean wavelength of the ISRF (see Equation \ref{urad}).

LH07 and HL08 show that RATs for irregular grains can be approximately described by an analytical model (AMO, see Appendix \ref{apdx:AMO}). The combination of Equation (\ref{eq:QRAT_avg}) with self-similar functional forms of RATs provided by AMO and a unique parameter $q^{\max}=Q_{e1}^{\max}/Q_{e2}^{\max}$ can describe RATs for arbitrary grain shape and radiation wavelength (see Appendix \ref{apdx:AMO}). { The value $q^{\max}$ is found to change with grain shape and radiation field. For the range $q^{\max}\sim 0.5-4$, we found that AMO provides a good agreement with numerical RATs of some irregular shapes (see Sec 4.5 in LH07). We refer our readers to LH07 for a complete discussion on the AMO.}

\section{MRAT Alignment in the deterministic case}\label{sec:numdet}
\subsection{Equations of Steady Motion}
The alignment of grains by RATs is studied by following the evolution of grain angular momentum in the phase space (DW97; HL08). In the deterministic case where random excitation is disregarded, the orientation of the grain angular momentum is governed by RATs, gaseous damping, and magnetic torques.

It is convenient to study grain orientation using spherical coordinates, which is completely determined by three variables: the angle $\xi$ between the angular momentum vector {\bf J} and the magnetic field direction {\bf B}, the Larmor precession angle $\phi $ of ${\bf J}$ around {\bf B} and the value of the angular momentum $J$ (see the right panel of Figure \ref{fig:RFs}). As shown in Equation (\ref{eq:tauB}), the Larmor precession rate is increased by $N_{\rm cl}$, ensuring that the Larmor precession rate is much larger than $\tau_{m}^{-1}$ and $\tau_{\gas}^{-1}$. Therefore, the averaging of RATs over the Larmor precession is justified. 

In this section, we first disregard the thermal fluctuations within the grain and assume a perfect coupling of the axis of major inertia with $\bJ$ (e.g., LH07). Thus, the equations of motion for $J'=J/I_{\|}\omega_{T}$ and $t'=t/\tau_{\gas}$ become:
\bea
\frac{d\xi}{dt'}&=&\frac{M\bar{F}(\xi, \psi)}{J'} - \delta_{m}\sin\xi\cos\xi ,\label{eq:dxidt}\\ 
\frac{dJ'}{dt'}&=&M \bar{\H}(\xi,\psi)-J'\left(1 + \delta_{m}\sin^{2}\xi\right),\label{eq:dJdt}
\ena 
where $\omega_{T}=(2kT_{\gas}/I_{\|})^{1/2}$, $M={\gamma \bar{\lambda}u_{\rad} a}/{2}$, $\bar{F}(\xi,\psi)$ and $\bar{H}(\xi,\psi)$ are aligning and spin-up torque components averaged over the precession angle $\psi$ (see Appendix \ref{apdx:AMO}). 
          
\subsection{Conditions for High-J Attractor Points}
Let ($\xi_{s},J_{s}$) be stationary points satisfying equations $d\xi'/dt=0, dJ'/dt=0$. Possible solutions of Equations (\ref{eq:dxidt}) and (\ref{eq:dJdt}) are given by:
\bea
\frac{\bar{F}(\xi_{s})(1+\delta_{m}\sin^{2}\xi_{s})}{\bar{H}(\xi_{s})} -\delta_{m}\sin\xi_{s}\cos\xi_{s}=0.\label{eq:ffunc}
\ena

As shown in LH07,  when averaged over the fast Larmor precession, we have $\bar{F}=0$ at $\sin\xi_{s}=0$. Therefore, Equation (\ref{eq:ffunc}) reveals that $\sin\xi_{s}=0$ are universal stationary points. In addition, there exists intermediate stationary points $\xi_{s}=(0,\pi)$, depending on $\delta_{m}$ and RATs. 

Following DW97, the stationary point $(\xi_{s}, J_{s})$ is an attractor if
\bea
A+D<0,~ BC-AD<0,\label{eq:criattractor}
\ena
where
\bea
A&=&\frac{M}{J'_{s}}\frac{d\bar{F}}{d\xi} -\delta_{m}\cos 2\xi_{s},~ B=-\frac{M}{J_{s}^{'2}}\bar{F}(\xi_{s}),\\
C&=& M\frac{d\bar{H}}{d\xi} -\delta_{m} J'_{s}\sin 2\xi_{s},~ D=-\left(1+\delta_{m} \sin^{2}\xi_{s}\right).~~\label{eq:attrpoint}
\ena

For the universal stationary point $\sin\xi_{s}=0, J'_{s}=M\bar{H}(\xi_{s})$, the condition for the attractor point (Eq. \ref{eq:criattractor}) is reduced to a single inequality
\bea
\left. \frac{1}{\bar{H}}\frac{d\bar{F}}{d\xi}\right|_{\sin\xi_{s}=0} -\delta_{m}<0,\label{eq:attractor}
\ena
which simply indicates that to produce an attractor point, the rate of increasing $\xi$ from $\xi_{s}$ by RATs must be lower than the rate of decreasing $\xi$ by magnetic relaxation $\delta_{m}$. { It can be seen that the condition of high-J attractor does not depend on the anisotropy degree of radiation $\gamma$ provided that $\gamma\ne 0$.}
 
The discussion above is applied for helical grains of right helicity, i.e., when the radiation is parallel to the field, the spin-up torque $\bar{H}$ is positive (LH07). It is applied also for left helicity in which the high-J attractor points is $\xi_{s}=\pi, \bar{H}(\xi_{s})>0$.

\subsection{Results for high-J attractor points using AMO}
In a special case when the anisotropic radiation direction is parallel to the magnetic field (i.e., $\psi=0$), using RATs from AMO (see Appendix \ref{apdx:AMO}), LH08 found that the new high-J attractor point is present when $\delta_{m}$ satisfies the following condition:
\bea
\delta_{m}>\delta_{\rm m, cri}=\frac{2-q^{\max}}{q^{\max}}.\label{eq:deltam_cri}
\ena
The above relation indicates that, in the case of $\psi=0^{\circ}$, the high-J attractor is always present even for ordinary paramagnetism if $q^{\max}>2$ (LH08).

To derive a parameter space for attractor points for arbitrary angle $\psi$, we employ RATs from AMO and solve Equation (\ref{eq:attractor}) numerically. 

\begin{figure*}
\centering
\includegraphics[width=0.6\textwidth]{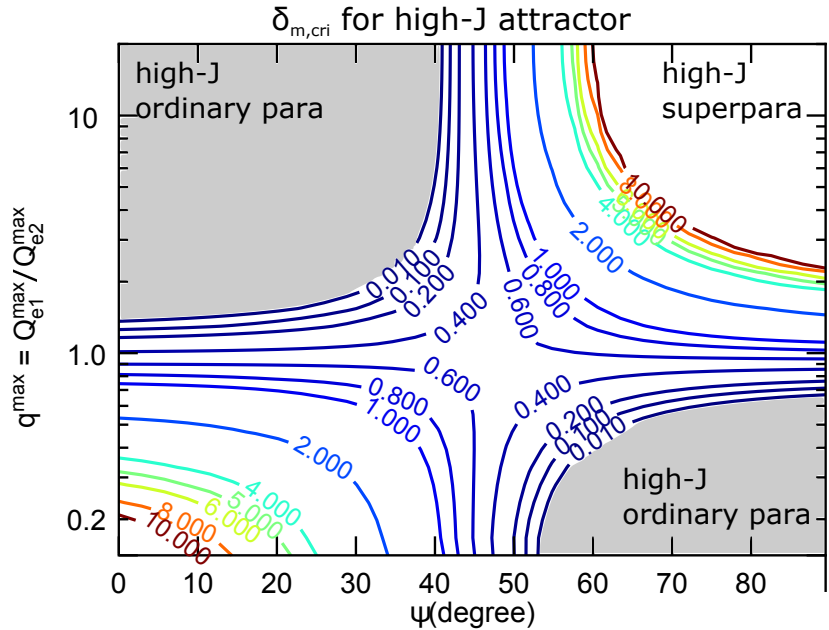}
\caption{Contours of the critical value of magnetic relaxation parameter required to produce high-J attractor points, $\delta_{\rm m, cri}$, in the plane of $\psi$ and $q^{\max}$. Shaded areas indicate the existence of high-J attractor points produced by RATs even for ordinary paramagnetic grains (i.e., no need of magnetic relaxation). White areas indicate the space where the high-J attractor points are only produced in the presence of magnetic relaxation. High-J attractors is universal for $\delta_{\rm m,cri}>10$.}
\label{fig:deltam_cri}
\end{figure*}

Figure \ref{fig:deltam_cri} shows the contours of $\delta_{\rm m,cri}$ required to produce high-J attractor points in the plane of $\psi$ and $q^{\max}$. For a wide range of explored parameters of $q^{\max}$ and $\psi$, high-J attractor points are present for $\delta_{\rm m,cri}\le 10$, except for the case $\psi=45^{\circ}$. For instance, for $\psi=0$, higher $\delta_{m}$ is required to have high-J attractors for smaller $q^{\max}$, consistent with the analytical relation shown in Equation (\ref{eq:deltam_cri}). It is noted that the condition in Figure \ref{fig:deltam_cri} is applicable to any value of $a$. { Moreover, the result is independent on the anisotropy $\gamma$, provided that $\gamma\ne 0$.}

\subsection{Physical Parameters}
{ For numerical simulations in the following, we adopt the typical physical parameters for the average interstellar radiation field (ISRF) in the solar neighborhood (\citealt{1983A&A...128..212M}) with the energy radiation density of $u_{\rm ISRF}=8.64\times 10^{-13}\erg\cm^{-3}$ (see Equation \ref{urad}) and anisotropy degree $\gamma=0.1$ (DW97). We also consider the typical CNM for the ISM with $n_{\H}=30\cm^{-3}$, $T_{\gas}=100$K, $T_{d}=20$K, and $B=10\mu$G (see Table \ref{tab:notation}).} 

\subsection{Trajectory maps in the steady rotation regime}
To visualize grain alignment by RATs, it is convenient to follow the temporal evolution of grain orientation from some initial orientations and construct the trajectory map of grain orientation. { We consider an ensemble of $N_{\rm gr}=21$ grains with initial angular momentum $J_{0}=50I_{1}\omega_{T}$ and initial orientations given by a regular grid from $\xi_{0}=0-\pi$.} Then, we solve Equations (\ref{eq:dxidt}) and (\ref{eq:dJdt}) for $\xi(t)$ and $J(t)$ with timestep $dt'=\min[5\times 10^{-4}, 10^{-3}/\delta_{\rm m}]$. With this choice of $dt'$, one has $dt'\gg \tau_{\rm Lar}$ (Eq \ref{eq:tauB}), justifying the averaging of the equations of motion over the fast Larmor precession. 

\begin{figure*}
\centering
\includegraphics[width=0.7\textwidth]{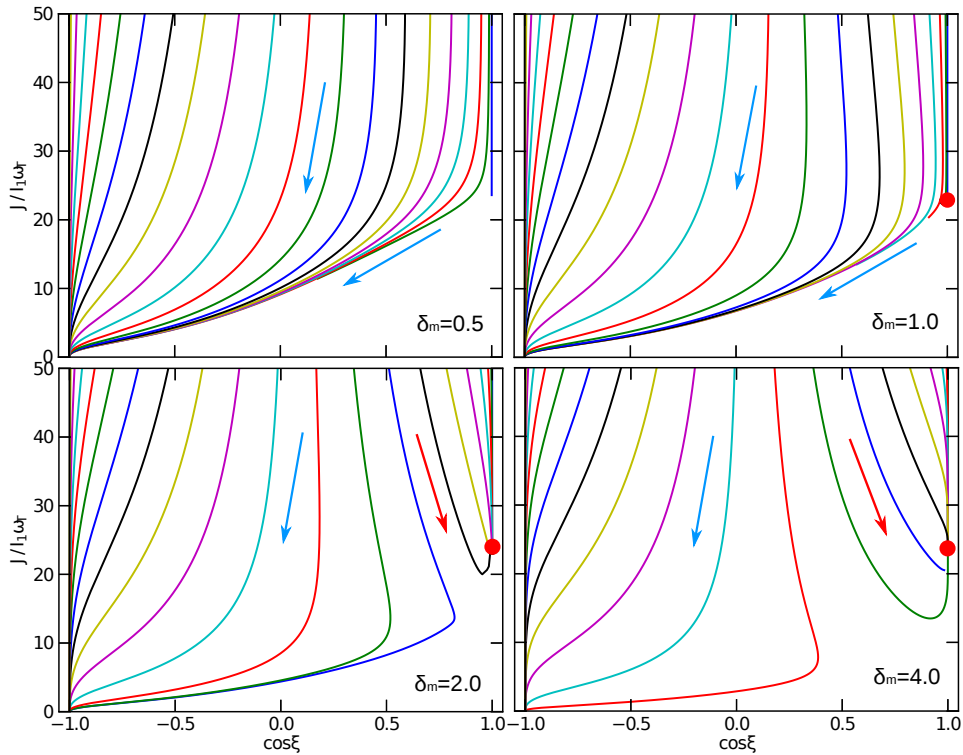}
\caption{Phase trajectory maps of MRAT alignment for $q^{\max}=1$ and $\psi=0^{\circ}$. Four different values of $\delta_{\rm m}$ are considered. Red arrows represents grains heading to the high-J attractor point (red filled circle), while blue arrows represent grains proceeding toward the low-J rotation. The high-J attractor points appear as $\delta_{\rm m}$ increases from $0.5$ to $2$, consistent with the condition shown in Figure \ref{fig:deltam_cri}.}
\label{fig:map_ncoll}
\end{figure*}

Figure \ref{fig:map_ncoll} shows the trajectory maps for the selected cases of $q^{\max}=1$ and four different values of magnetic relaxation parameters $\delta_{\rm m}=0.5, 1, 2, 4$. { The effective grain size $a=0.1\mum$ is assumed}. As shown, grains are driven to low-J and high-J attractor points, regardless of their initial states. Moreover, the high-J attractor points are created when $\delta_{\rm m}$ is increased as predicted. For instance, the alignment does not have high-J attractor point for the low value of $\delta_{\rm m}=0.5$, but the high-J attractor point appears when increasing the magnetic relaxation to $\delta_{\rm m}= 2$, as predicted in Figure \ref{fig:deltam_cri}. 

As shown, most of grains are driven to low-J attractor points. However, during the period of non-suprathermal rotation (e.g., $J/I_{1}\omega_{T}< 5$), randomization by gas collisions should be taken into account, and the dynamics of grains will be dramatically different, as demonstrated in the next section.

\section{MRAT alignment in the presence of stochastic excitations}\label{sec:numstoc}
For grains aligned at low-J attractor points rotating at thermal speeds, HL08 pointed out that random gas collisions can have counter-intuitive effect of enhancing the alignment. In this section, we first describe numerical method that treat both random excitations and MRAT alignment. Then, we will carry out extensive calculations for different values of $\delta_{\rm m}$, $q^{\max}$, $\psi$, and grain size $a$. { Moreover, to obtain realistic measurements of MRAT alignment, we now also account for the effect of thermal fluctuations within the grain \citep{1997ApJ...484..230L} on the RATs (HL08). Thermal fluctuations are important for ordinary paramagnetic grains smaller than $\sim 1\mum$ (HL09b), and superparamagnetic inclusions can significantly increase this range (\citealt{Lazarian:2008fw}).} Therefore, we first average radiative torque components $F$ and $H$ over thermal fluctuations and then over the precession to obtain tabulated data for $\langle {F}(\xi, J)\rangle$ and $\langle{H}(\xi, J)\rangle$ as functions of $\xi, J$. These averaged torques will be plugged into Equations (\ref{eq:dxidt}) and (\ref{eq:dJdt}) to solve for $J$ and $\xi$ (see HL08 for details).

\subsection{Equations of Stochastic Motion: Langevin Equations}
The effect of gas collisions in the framework of paramagnetic alignment was studied 
by many authors (JS67; \citealt{1971ApJ...167...31P};\citealt{1997MNRAS.288..609L}) using the Fokker-Planck equations. The Langevin equation (LE) approach was used to study this problem numerically in \cite{1993ApJ...418..287R} (hereafter R93) and \cite{1999MNRAS.305..615R}. It is recently applied to study the effect of collisional excitations in the framework of RAT alignment (HL08), spinning dust emission (\citealt{Hoang:2010jy}; \citealt{2014ApJ...790....6H}), and magnetic dust emission (HL16). 

Let ($\xhat,\yhat,\zhat$) be the inertial frame of reference where $\zhat$ is defined parallel to the magnetic field. An increment of grain angular momentum after time interval $dt$ due to various random interactions can be described by the LE (\citealt{1993ApJ...418..287R}) 
\bea 
d J_{i}=A_{i}(t)dt+B_{ij}(J,t)dw_{j},~ i=x,y,z,\label{eq:LE}
\ena
where $d\omega_{j}$ are Wiener coefficients, and $A_{i}, B_{ij}$ are diffusion coefficients defined as 
\bea
A_{i}&=\langle \Delta J_{i} \rangle,~ i=x,y,z, \label{eq:Acoef}\\ 
(BB^{T})_{ij}&=\langle \Delta J_{i} \Delta J_{j}\rangle,i,j=x,y,z,\label{eq:Bcoef}
\ena
where $B^{T}$ is the transposal matrix of $B$ (see Appendix \ref{apdx:collexc}). 

{As in HL08, to account for the randomization effect by gas collisions, we employ the hybrid approach.} For each time step, we first solve the LEs subject to gas-grain collisions and magnetic fluctuations by solving for three components $J_{x}, J_{y}, J_{z}$ using the second-order integrator implemented in \cite{2016ApJ...821...91H} (see also Appendix \ref{apdx:collexc}), and compute $\tilde{J}=\sqrt{J_{x}^{2}+J_{y}^{2}+J_{z}^{2}}$, $\cos\tilde{\xi}=J_{z}/\tilde{J}$. Then, we use $\tilde{J}$ and $\tilde{\xi}$ as input parameters to solve Equations (\ref{eq:dxidt}) and (\ref{eq:dJdt}) for new values of $\xi, J$, where the second terms involving the gas and magnetic damping are removed because those effects are already accounted for by the LEs. { The advantage of using the hybrid approach is that it avoids the problem with some negative value of $J$ which may be generated in Monte Carlo simulations when directly solving $J,\xi$ in the spherical coordinate system using Equations (\ref{eq:dxidt}) and (\ref{eq:dJdt})}. For estimates of rotational damping and excitation coefficients by gas collisions and magnetic fluctuations, we adopt oblate spheroidal shape for grains (see Section \ref{apdx:collexc}).

\subsection{Simulation Setup}

{ We perform numerical simulations for a wide range of $\delta_{\rm m}=0.5-30$. This range of $\delta_{\rm m}$ covers a reasonable parameter space of magnetic inclusions which is easily achieved with less than $15\%$ of Fe incorporated in the form of iron clusters. Observations reveal that a significant fraction of Fe is locked in the dust \citep{2009ApJ...700.1299J}, so we expect to have much larger $\delta_{m}$. However, as shown in Figure \ref{fig:deltam_cri}, at $\delta_{m}\sim 10$, the high-J attractor point is already present for the selected parameter space. Therefore, the upper value $\delta_{m}=30$ is sufficient to capture the effect of iron inclusions on grain alignment. 
We consider four realizations of RATs from AMO with $q^{\max}=0.5,1,2,4$. This range of $q^{\max}$ covers the  realistic range of $q^{\max}$ that is found for numerical RATs computed for irregular grains in the range $\lambda/a=0.1-10$ with dominant RATs (see LH07). For the initial value of grain angular momentum, we take $J_{0}=3J_{\th}$ for $a=0.05\mum$ and $J=25J_{\th}$ for $a\ge 0.08\mum$ because the latter large grains are expected to have suprathermal rotation by RATs (HL09a).}

To check if grains converge to some attractor point, we introduce the convergence criteria as follows:
\bea
\frac{|J_{i+1}-J_{i}|}{J_{i}} \le \epsilon_{J},\
\frac{|\cos(\xi_{i+1})-\cos(\xi_{i})|}{|\cos(\xi_{i})|} \le \epsilon_{\xi},
\ena
where $i$ is the timestep, and we adopt $\epsilon_{J}=10^{-6}$ and $\epsilon_{\xi}=10^{-6}$.

\subsection{Trajectory maps}
\begin{figure*}
\centering
\includegraphics[width=0.7\textwidth]{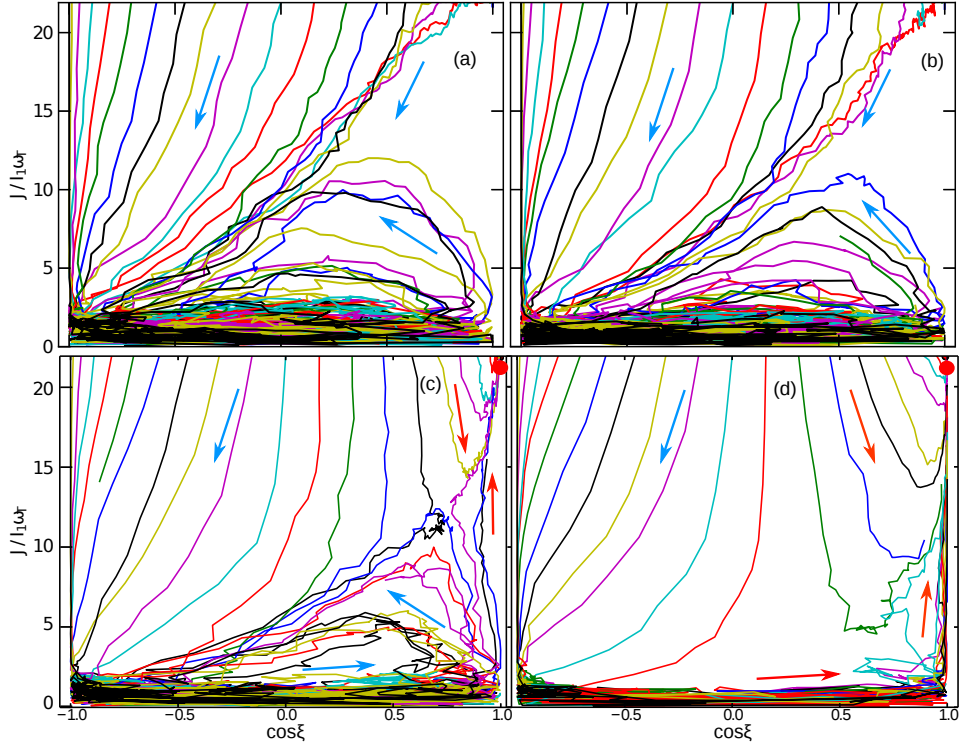}
\caption{Same as Figure \ref{fig:map_ncoll}, but for the case stochastic excitations by collisions and magnetic fluctuations are considered. Red arrows represents grains heading to the high-J attractor point (red circles), while blue arrows represent grains follow cyclic trajectories. In the presence of the high-J attractor, grains at the low-J attractor are strongly disturbed and excited to the high-J attractor (see (c) and (d)), whereas grains follows perturbed cyclic trajectories in the absence of high-J attractors (see (a) and (b) panels).}
\label{fig:map_coll}
\end{figure*}

Figure \ref{fig:map_coll} shows the trajectory maps for four cases, without high-J attractors ((a) and (b)) and with high-J attractors ((c) and (d)). Same as Figure \ref{fig:map_ncoll}, but random excitations important during the slow rotation period are treated. 

Let us discuss first the trajectory maps without high-J attractors (panels (a) and (b)). It can be seen that random collisions have little effect on the grain trajectory during the initial suprathermal rotation period (i.e., $J/I_{1}\omega_{T}\gg 1$). After some time, grains are driven to the thermal rotation stage ($J/I_{1}\omega_{T}< 3$) during which stochastic excitations by collisions can have dramatic effects. Indeed, grains are no longer being aligned at the low-J attractor (cf. (a) and (b) in Figure \ref{fig:map_ncoll}), instead, grains are scattered in the entire phase space until RATs become active again in driving grains back to the vicinity of $\cos\xi=-1$. The trajectory map can be considered as perturbed cyclic trajectories. 

For the case with high-J attractor points (panels (c) and (d)), grains at the low-J attractor point are scattered randomly in its phase space. At some point, grains are scattered into the vicinity of $\cos\xi\sim 1$, for which the spin-up component of RATs rapidly drives grains to the high-J attractor point. 

\subsection{Time-Dependence Degree of Alignment and Averaged Degree of Alignment}
{ In addition to the grain shape and its composition, the degree of grain alignment is an important parameter for modeling the polarization spectrum of thermal dust emission as well as MDE}. In this section, we will carry simulations to calculate the degree of MRAT alignment for a broad range of relevant parameters.

To quantify the time-dependent degree of grain alignment, we compute the ensemble average of the alignment of the angular momentum with the magnetic field, $\tilde{Q}_{J}$, and the degree of alignment of the grain axis with $\Bv$ (i.e., Rayleigh reduction factor; \citealt{Greenberg:1968p6020}), $\tilde{R}$, at each timestep as follows:
\bea
\tilde{Q}_{J}=\frac{1}{N_{\rm gr}}\sum_{i=1}^{N_{\rm gr}} \frac{(3\cos^{2}\xi_{i}-1)}{2},\label{eq:QJ}\\
\tilde{R}=\frac{1}{N_{\rm gr}}\sum_{i=1}^{N_{\rm gr}} \frac{(3\cos^{2}\xi_{i}-1)\tilde{Q}_{X}(J_{i})}{2},
\ena
where $N_{\rm gr}$ is the number of grains in our simulations, and $\xi_{i}$ is the value of $\xi$ for grain $i$ at the end of each integration step. The value of $\tilde{Q}_{X}$ (i.e., the degree of alignment of grain axis with $\bJ$) is obtained by integrating over the fast internal fluctuations as follows (see \citealt{1997ApJ...484..230L} for detail):
\bea
\tilde{Q}_{X}(J_{i})\propto \int_{0}^{\pi} q_{X}\exp\left({-J_{i}^{2}[1+(h-1)\sin^{2}\theta]}\right)\sin\theta d\theta,~~~
\ena
where $h=I_{\|}/I_{\perp}$, and $q_{X}=(3\cos^{2}\theta-1)/2$ (see Appendix \ref{apdx:collexc}).

{ We have simulated for three values of $N_{\rm gr}=16, 21$, and 32 and found that the results are within $1\%$ difference. Therefore, we chose $N_{\rm gr}=21$ for our results presented in this paper.}

To compute the average degree of grain alignment, we make use of ergodicity of grain dynamical system for which the average of the degree of alignment over an ensemble of grains can be replaced by the time average. In that sense, the value of $R$ is computed as follows:
\bea
R=\frac{1}{N_{T}}\sum_{i=N_{0}}^{N_{T}}\tilde{R}(t_{i}),\label{eq:Ravg}
\ena
where { $N_{T}$ is the total number of integration steps}, and $N_{0}$ is the timestep from the moment long enough so that grains forget their initial states. ${Q}_{X}$ and ${Q}_{J}$ are computed as $R$ as given by Equation (\ref{eq:Ravg}).

{ The value $N_{T}$ is determined by $T/dt$ where $T$ is the integration time and $dt$ is the timestep. To ensure a good statistic on alignment measurements, we integrate over a long timescale $T=200\tau_{\gas}$. 
The value of $N_{0}$ is chosen at the timestep so that the averaging time interval $\Delta t_{\rm avg}=(N_{T}-N_{0})dt =50\tau_{\gas}$. We have checked with $T=300\tau_{\gas}$ and $400\tau_{\gas}$ and find that the results are essentially the same. We have tested with different $N_{0}$ given by $\Delta t_{\rm avg}=10\tau_{\gas}-50\tau_{\gas}$ and found that the results differ by only less than $3\%$ for a fixed $N_{T}$.}

\begin{figure*}
\centering
\includegraphics[width=0.7\textwidth]{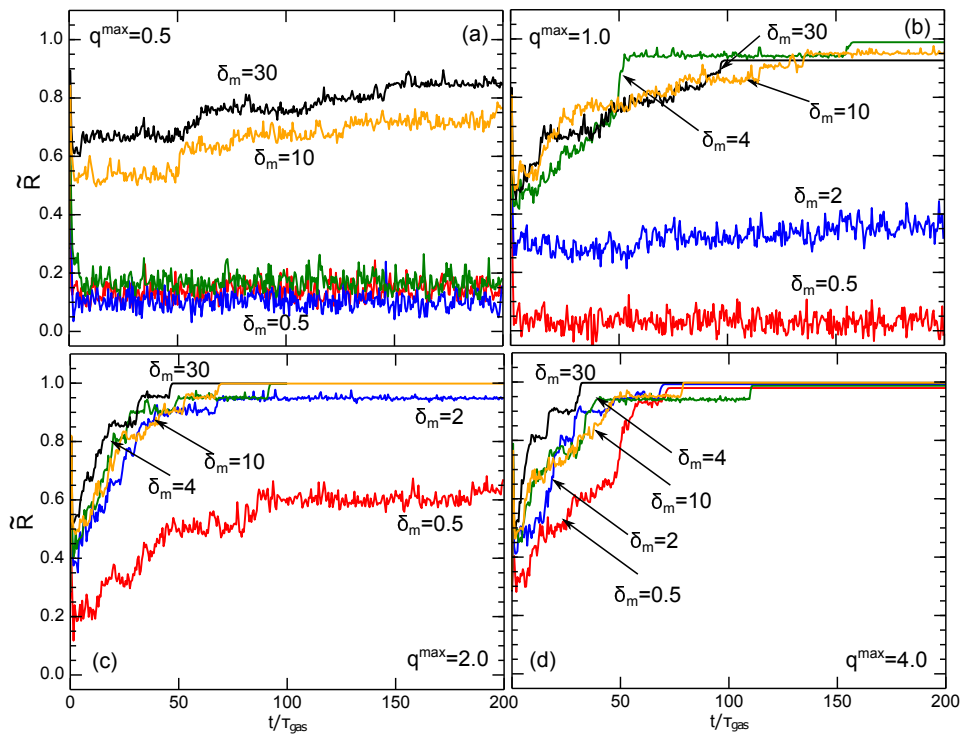}
\caption{Variation of $\tilde{R}$ with $t/t_{\gas}$ for $q^{\max}=0.5$ (a), 1 (b), 2(c), and $4$ (d). Different values $\delta_{\rm m}=0.5, 2, 4, 10, 30$ are indicated in each panel. Perfect alignment is achieved for the largest value of $\delta_{\rm m}$, except the first case (panel (a)). The grain size $a=0.1\mum$ and anisotropic radiation direction $\psi=0^{\circ}$ are assumed.}
\label{fig:Rtime2_coll}
\end{figure*}

Figure \ref{fig:Rtime2_coll} shows the time-dependence value of $\tilde{R}$ for four different values of $\delta_{\rm m}$ and $\psi=0-80$ degree. For the case with high-J attractor points, we see the gradual increase of alignment degree with time and reach perfect alignment after $t\sim 40\tau_{\gas}$. Second, the time of perfect alignment is smaller for higher $\delta_{\rm m}$. In addition to driving some grains to high-J attractor points, as discovered in HL08, we found that collisional excitations result in the increase of the rms angular momentum of grains, increasing the internal alignment $\tilde{Q}_{X}$. For $\delta_{\rm m}=10$, perfect alignment is achieved regardless of $q^{\max}$. Without high-J attractor points (green and purple lines in (a)), we can see that the degree of alignment is still moderate, at level of $\tilde{R}\sim 0.2$. This is because, although following disturbed trajectories, the grains spend a large fraction of time in the vicinity of two stationary points of $\cos\xi=\pm1$ of perfect alignment.

\section{Parameter space studies for the MRAT alignment}\label{sec:paraspace}

\subsection{Dependence on magnetic relaxation parameter $\delta_{\rm m}$}
We first study the dependence of the degree of alignment on $\delta_{\rm m}$.

\begin{figure*}
\centering
\includegraphics[width=0.7\textwidth]{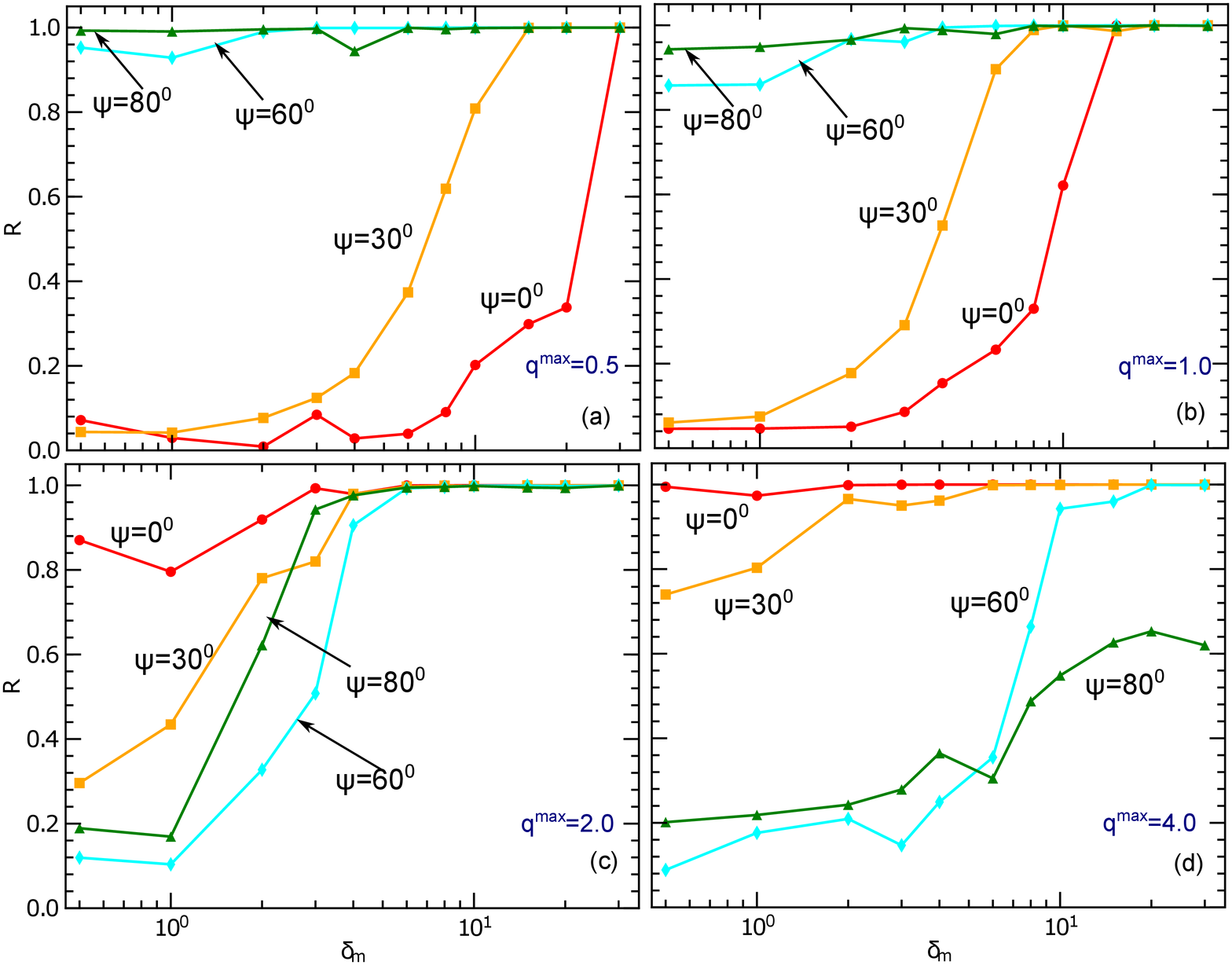}
\caption{Variation of the Rayleigh reduction factor $R$ with $\delta_{\rm m}$ for $\psi=0, 30, 60, 80^{\circ}$ and $q^{\max}=0.5, 1, 2, 4$ (panels (a)-(d)). The MRAT alignment is nearly perfect for the largest $\delta_{\rm m}$ where the alignment has high-J attractor points, such that all grains eventually become aligned. Grain size $a=0.2\mum$ is considered.}
\label{fig:Ravg_delta}
\end{figure*}

\begin{figure*}
\centering
\includegraphics[width=0.7\textwidth]{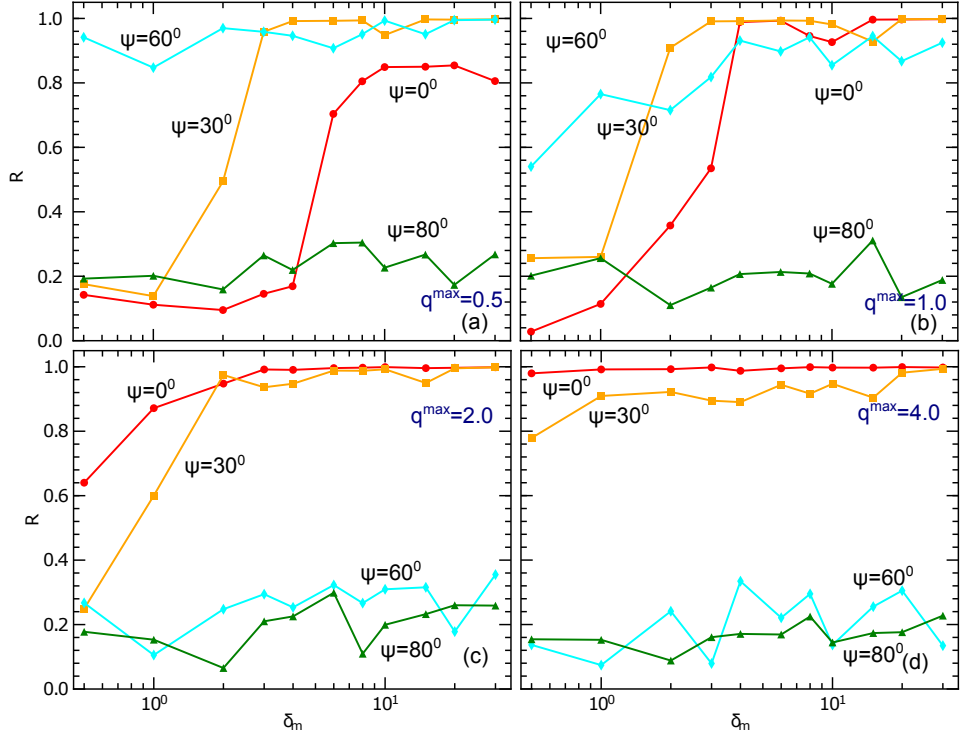}
\caption{Same as Figure \ref{fig:Ravg_delta}, but for $a=0.1\mu$m. Compared to Figure \ref{fig:Ravg_delta}, for the last two cases ((c) and (d)), the MRAT alignment cannot be perfect because these grains rotate at lower speeds than those of $a=0.2\mum$.}
\label{fig:Ravg_delta_a01}
\end{figure*}

\begin{figure}
\includegraphics[width=0.45\textwidth]{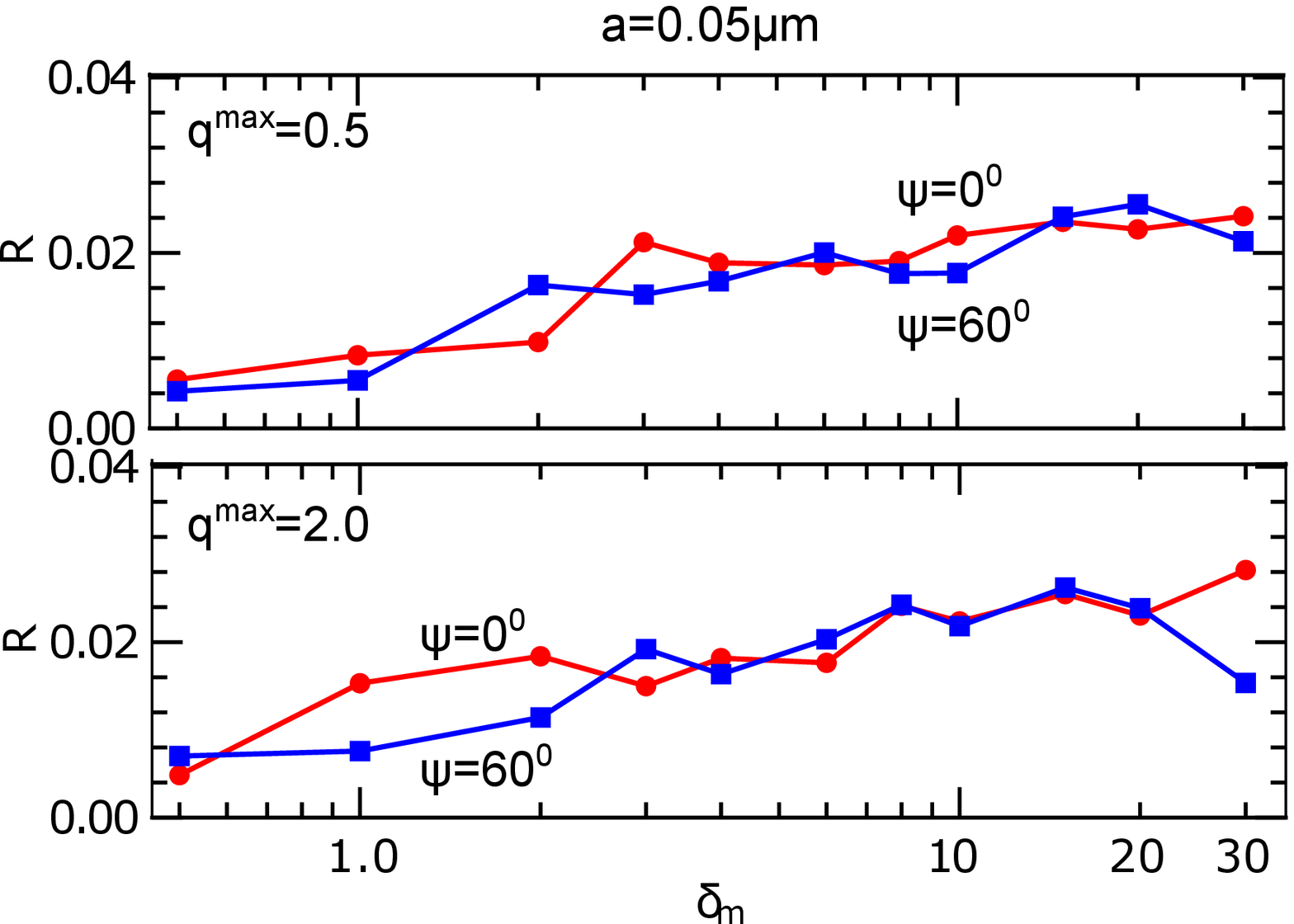}
\caption{Same as Figure \ref{fig:Ravg_delta}, but for $a=0.05\mu$m. Two angles $\psi=0^{\circ}$ and $60^{\circ}$ are shown.}
\label{fig:Ravg_delta_a005}
\end{figure}

Figure \ref{fig:Ravg_delta} shows the dependence of $\tilde{R}$ on $\delta_{\rm m}$ for the different values of $q^{\max}$ for grains of size $a=0.2\mum$. For $q^{\max}=0.5$ and $1$, the MRAT is almost perfect for $\psi=60^{\circ}$ and $80^{\circ}$ for all values of $\delta_{\rm m}$, whereas the perfect alignment is only achieved with large value of $\delta_{\rm m}$ for $\psi=0^{\circ}$ and $30^{\circ}$.
This arises directly from Figure \ref{fig:deltam_cri} that the high-J attractor points are present at $\psi\ge 60^{\circ}$ even with low values of $\delta_{\rm m}$. For $q^{\max}=2$ and $4$, the same happens with $\psi=0^{\circ}$ and $30^{\circ}$. 
For all cases of $q^{\max}$, the MRAT alignment becomes perfect for $\delta_{\rm m}\ge 30$ (panels (a)-(c)). For the case of $q^{\max}=4$, the value $\tilde{R}$ for $\psi=80^{\circ}$ cannot reach perfect alignment because grains do not rotate suprathermally at this large angle between the radiation and the magnetic field. 

In Figure \ref{fig:Ravg_delta_a01}, we present the obtained results for $a=0.1\mum$. The results are essentially similar to $a=0.2\mum$, but the value of $\delta_{\rm m}$ at which perfect alignment is achieved tends to be smaller. This is because these grains have weaker RATs, such that lower $\delta_{\rm m}$ are required to stabilize the high-J attractor points. It is also noted that grains with $\psi=80^{\circ}$ in all cases never reach perfect alignment because of thermally rotation. Compared to Figure \ref{fig:Ravg_delta}, for the last two cases ((c) and (d)), the MRAT alignment cannot be perfect because these grains rotate at lower rates than those of $a=0.2\mum$.

{ Figure \ref{fig:Ravg_delta_a005} shows the results for $a=0.05\mum$. The degree of alignment is rather small and is essentially independent of $\psi$. The value of $R$ tends to increase with $\delta_{m}$ and saturates for $\delta_{m}>20$. This is consistent with our results in HL16 for superparamagnetic grains rotating thermally.}

\subsection{Dependence on the anisotropic radiation direction $\psi$}

\begin{figure*}
\centering
\includegraphics[width=0.8\textwidth]{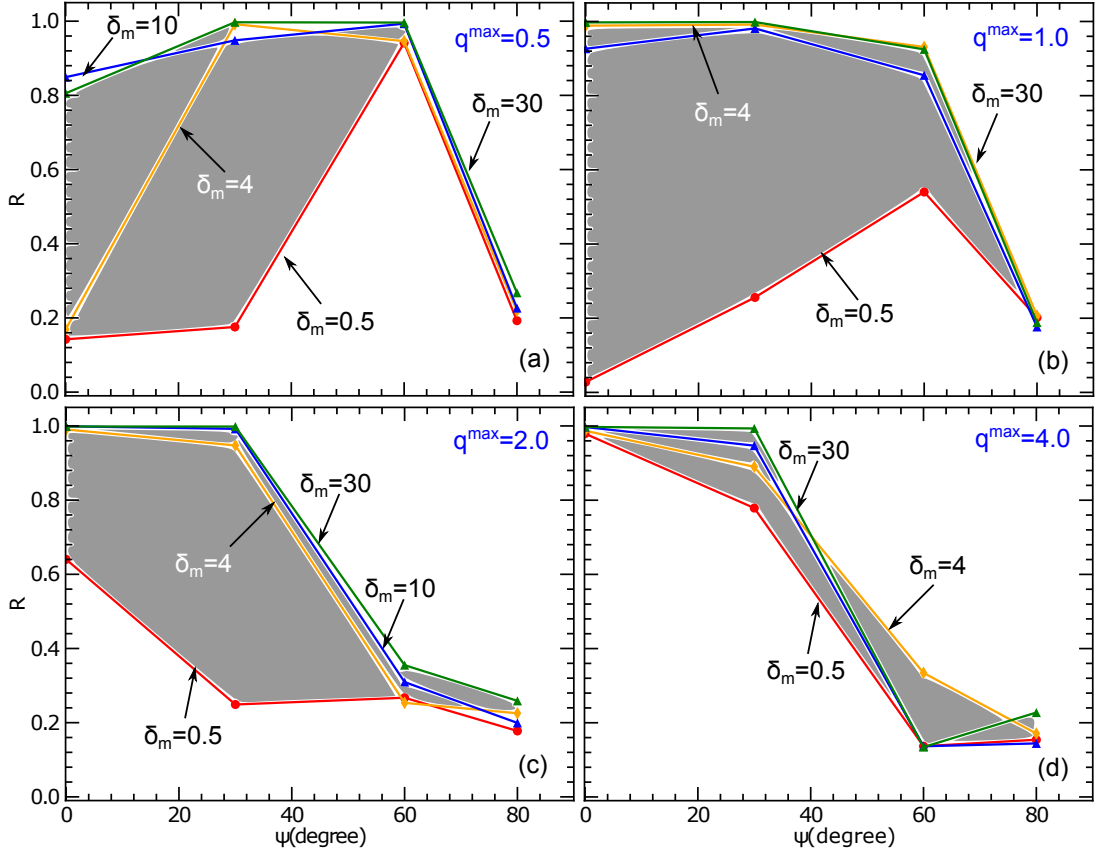}
\caption{Dependence of $R$ on the anisotropic radiation angle $\psi$ for different values of $q^{\max}$(panels (a)-(d)). Shade area indicate the range of $\delta_{\rm m}=0.5$ to $\delta_{\rm m}=30$. Selected values of $\delta_{\rm m}$ are indicated. A decrease of $R$ with increasing $\psi$ is found for $q^{\max}=2$ and $4$, while a more complex dependence is seen for $q^{\max}=0.5$ and $1$. The grain size $a=0.1\mu$m is considered.}
\label{fig:Ravg_psi}
\end{figure*}

Figure \ref{fig:Ravg_psi} shows an explicit dependence of the degree of alignment on $\psi$ for four values of $q^{\max}$.  For $q^{\max}=2$ and $4$, the value of $R$ tends to decrease with increasing $\psi$ for arbitrary $\delta_{\rm m}$. For $q^{\max}=1$, such a decreasing trend is observed for slightly enhanced susceptibilities with $\delta_{\rm m}>4$. The case of $q^{\max}=0.5$ exhibits a complicated 
dependence on $\psi$ even for large value of $\delta_{\rm m}\sim 30$.

\subsection{Dependence on grain size $a$}
\begin{figure*}
\centering
\includegraphics[width=0.8\textwidth]{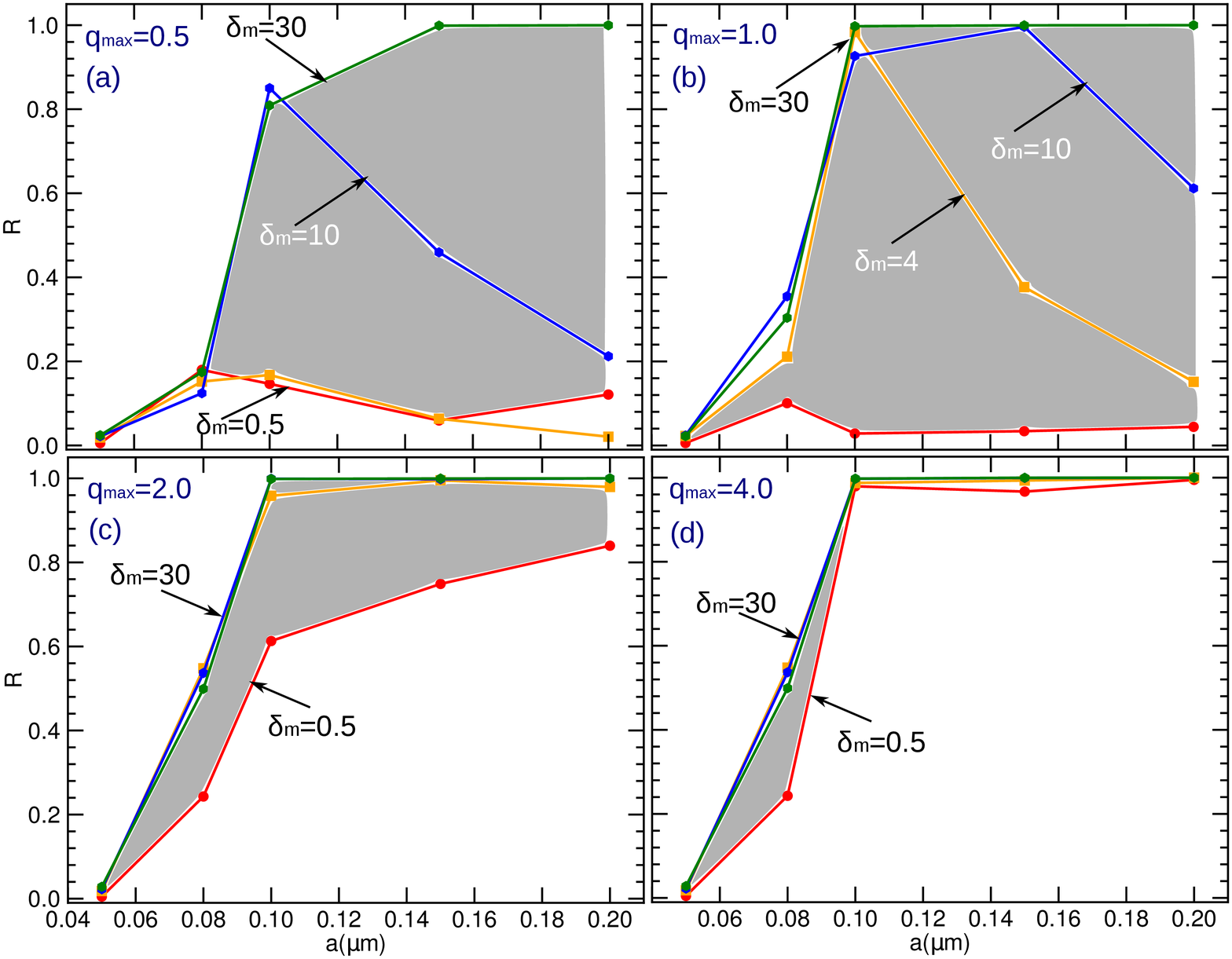}
\caption{Same as Figure \ref{fig:Ravg_psi}, but for the dependence of $R$ on the grain size $a$. Big grains ($a\ge 0.1\mum$) are mostly perfectly aligned for $\delta_{\rm m}\ge 30$, while grains $a<0.1\mum$ tend to increase with $a$. The results for $\psi=0^{\circ}$ are shown.}
\label{fig:Ravg_size}
\end{figure*}

Figure \ref{fig:Ravg_size} shows the dependence of $R$ on the grain size $a$ for different values of $q^{\max}$ at $\psi=0^{\circ}$. It can be seen that the values of $R$ increase with increasing $a$. For the largest magnetic considered, i.e, $\delta_{\rm m}=30$, the $a\ge 0.1\mum$ grains are perfectly aligned (upper bound) for all values of $q^{\max}$ considered, whereas smaller grains of $a=0.05$ and $0.08\mum$ are less efficiently aligned. 

For ordinary paramagnetic case, the values of $R$ are dependent on $q^{\max}$. For $a=0.05-0.08\mum$, we found that, even $\delta_{\rm m}\gg 1$, those small grains never reach perfect alignment. This can be explained by the fact that the latter smaller grains are not rotating suprathermally because of weaker RATs. Moreover, for the case of $\psi=60^{\circ}$, we do not have perfect alignment of $a=0.1\mum$ because those grains only rotate at thermal rates. Furthermore, the parameter space for the dependence of $R$ vs. $a$ is quite broad for the RATs with $q^{\max}=0.5$ and $1$, and it becomes very focused for the two other cases of $q^{\max}=2$ and $4$.

\section{Special cases of grain alignment}

\subsection{Alignment of very big grains}
{ Now let us evaluate the maximum size of dust grains that can be aligned with respect to the magnetic field in the light of our unified model of grain alignment.}

{ Dust grains in the diffuse interstellar medium follow some size distribution with most of their mass contained in the size range of $a<0.3\mum$ (\citealt{2001ApJ...548..296W}).} Grains in dense molecular clouds grow larger due to coagulation and accretion processes of gas atoms onto the ice mantle (\citealt{2001ApJ...548..296W}). Such grains can be easier aligned by reddened radiation in molecular clouds (LH07, Figure 32). This effect was involved in \cite{2005ApJ...631..361C} to explain the alignment of grains in molecular clouds without embedded stars (see also  \citealt{2007ApJ...663.1055B}). This allowed to explain the observations in \cite{2008ApJ...674..304W} and \cite{Jones:2014fk}, where polarization was observed from dark cores with $A_v>15$. Larger grains can be aligned by RATs in accretion disks as discussed first by \cite{2007ApJ...669.1085C}. There, however, the observations were not showing good correspondence with theoretical predictions (see \citealt{2009ApJ...704.1204H}). The results were also critically analyzed in \cite{2014MNRAS.438..680H}. Two reasons that are responsible for lower degrees of alignment of grains are the failure of the internal relaxation to bring the grain angular momentum ${\bf J}$ to align with the axis of the maximal moment of inertia $\hat{a}_{1}$, and the slow Larmor precession ${\bf J}$ about the direction of the ambient magnetic field. The former effect does not completely suppress the RAT alignment, as shown by \cite{2009ApJ...697.1316H}, but makes the alignment more selective in terms of the grains that can be aligned. The effect of the reduced Larmor precession destroys the alignment with respect to the magnetic field. 

In view of the above, the effect of MRAT alignment is not only to allow better alignment due to stabilizing the high-$J$ attractor points, but also extending the range of grain sizes for which grains experience efficient internal relaxation (see LH07) and faster Larmor precession. In the presence of strongly magnetic inclusions, the magnetic moment due to Barnett effect is stronger (see Equation (\ref{eq:muBar})):
\bea
\bmu_{\Bar,sup}=-\frac{\chi_{\rm sp}(0)\hbar V}{g_{e}\mu_{B}}\bomega,\label{eq:muBar_sup}
\ena
which induces internal relaxation with characteristic timescale
\bea
\tau_{\rm Bar,sup}=\frac{\gamma_{e}^{2}I_{\|}^{3}}{VK(\omega)h^{2}(h-1)J^{2}},\label{eq:tauBar}
\ena
where $h=I_{\|}/I_{\perp}$ (see Equation \ref{eq:Iparperp}), and $\gamma_{e}=g_{e}e/2m_{e}c$ is the magnetogyric ratio.

Using the Larmor precession time from Equation (\ref{eq:tauB}), one obtains the ratio of the Larmor precession to gas damping time as following:
\bea
\frac{\tau_{\rm Lar,sup}}{\tau_{\gas}}&=&\left(\frac{\tau_{\rm Lar,sup}}{\tau_{\rm mag}}\right)\left(\frac{\tau_{\rm mag}}{\tau_{\gas}}\right)= \frac{2\pi g_{e}\mu_{B}}{\hbar}\left(\frac{n_{\H}m_{\H}v_{\rm th}\Gamma_{\|}a_{2}}{\chi_{\sp}(0)sB}\right),\nonumber\\
&\simeq& 22.2\times 10^{-3} \frac{n_{10}T_{2}^{1/2}a_{-5}\Gamma_{\|}}{N_{\rm cl,5}\phi_{sp,-2}B_{3}},
\ena
where $n_{10}=n_{\H}/10^{10}\cm^{-3}$, $N_{cl,5}=N_{\rm cl}/10^{5}$, $\phi_{sp,-2}=\phi_{sp}/0.01$, and $B_{3}=B/1000\mu$G.

Therefore, the largest grain size that the Larmor precession is still important is determined by
\bea
a_{1} < 436 \frac{N_{cl,5}\phi_{sp,-2}}{n_{10}T_{2}^{1/2}B_{3}\Gamma_{\|}} \mum,\label{eq:aLar}
\ena
where $N_{\rm cl}$ spans $\sim 20$ to $10^{5}$ (JS67). For the upper estimate of $a_{1}$, we take $N_{cl}=10^{5}$.

{ For accretion disks with $B\sim 10$mG \citep{2006Sci...313..812G} and $T_{\gas}\sim 25$K \citep{1997ApJ...490..368C}, we expect to have alignment of very big grains with the magnetic field in the disk interior for $a_{1}\sim 218 N_{cl,5}/n_{10}\mum$. For a typical flared disk around T-Tauri stars, the density at radius $d$ is $n_{\H}\sim 10^{9}(d/100\AU)^{-39/14}\exp\left(-z^{2}/2H^{2}\right)\cm^{-3}$ with $H$ the scale height and $z$ the distance from the disk plane ( \citealt{2014MNRAS.438..680H} ). Therefore, grains as large as $a_{1}\sim 2$mm can be aligned in the diskplane at distance $d>100\AU$. At disk radius $d=50\AU$, grains upto $a_{1}\sim 0.5$mm can be aligned with the magnetic field in the disk plane; at height $z\sim 2H$, this upper size increases to $a_{1}\sim 4$mm. }

We move on to estimate the size at which internal relaxation is still important. From Equations (\ref{eq:tgas}) and (\ref{eq:tauBar}), it is straightforward to obtain:
\bea
\frac{\tau_{\rm Bar,sup}}{\tau_{\gas}}&=& \frac{\gamma_{e}^{2}n_{\H}mv_{\rm th}a\Gamma_{\|}}{s^{4/3}h^{2}(h-1)\omega^{2}}\times \frac{[(1+(\omega\tau_{sp}/2)^{2}]^{2}}{\chi_{\rm sp}(0)\tau_{\rm sp}},\\
&=& 8.5\times 10^{-5}\frac{n_{10}a_{-5}\Gamma_{\|}}{N_{cl,5}\phi_{\rm sp,-2}}\frac{[(1+(\omega\tau_{sp}/2)^{2}]^{2}}{\exp\left({N_{\rm cl}T_{\rm act}/T_{d}}\right)},
\ena
where $h=2$ is taken for the estimate.

Thus, the condition for potential internal alignment is satisfied for grains of size
\bea
a_{2} &<& 0.48 \left(\frac{n_{10}T_{2}^{-1/2}}{N_{cl,5}\phi_{sp,-2}}\right)^{1/6}\left(\frac{[(1+(\omega\tau_{sp}/2)^{2}]^{2}}{\exp\left({N_{\rm cl}T_{\rm act}/T_{d}}\right)}\right)^{-1/6}\nonumber\\
&&\times \left(\frac{\omega}{\omega_{T}}\right)^{1/3}~\mum.\label{eq:aBar}
\ena

{ For the similar conditions of T-Tauri disks aforementioned, we expect Barnett relaxation is important only for small grains upto $a_{2}\sim 0.5\mum$ at radius $d>100\AU$. Nevertheless, grains with inefficient internal relaxation may still be aligned with the magnetic field by strong RATs (HL09b) and that mechanical alignment perhaps plays some role in the disk. Therefore, the potential of alignment of the millimeter-sized grains with the magnetic field in the accretion disks remains feasible if grains are incorporated with iron inclusions. Testing the alignment in accretion disks can therefore be a good test of the MRAT mechanism.}

\subsection{Alignment of carbonaceous grains}

One related problem is the alignment of carbonaceous grains. Observations testify that these grains are not aligned (\citealt{2006ApJ...651..268C}), although RATs acting on these grains are not very different that those acting on silicate grains. In \cite{LAH15}, we discussed a possibility that carbonaceous grains may have a much smaller magnetic moment. Indeed, for the magnetic moment of carbonaceous grains can arise from the rotation of charged grains (\citealt{1971MNRAS.153..279M}). For a grain of mean charge $e\langle Z\rangle$ localized at distance $\epsilon a_{\perp}$ from the center of mass, the magnetic moment is
\bea
\mu_{\rm rot}&=&\frac{\mathcal{I}S }{c}=\frac{e\langle Z\rangle (\epsilon a_{\perp})^{2} \omega}{2c},\\
&\simeq& 1.6\times 10^{-13}a_{-5}^{2}(\epsilon/0.1)^{2}(\langle Z\rangle/30) \omega \D,
\ena
where $\mathcal{I}=e\langle Z\rangle \omega/2\pi$ is the electric current arising from charge rotation, and $S=\pi a_{\perp}^{2}=2\pi/3a^{2}$ is the surface area perpendicular to the current. 

In the typical interstellar magnetic field of $5\mu$G, this produces the Larmor precession of timescale:
\bea
\tau_{\rm Lar,rot}\sim 4\times 10^{4} a_{-5}^{3}(\epsilon/0.1)^{2}(\langle Z\rangle/30)~ \yr.
\ena

The magnetization of carbonaceous grains perhaps also originates from the attachment of H atoms to the grain surface through hydrogenation. Since a H electron is already used to make a covalent bond with a C atom, the magnetization is only produced by the H nucleus. The susceptibility is estimated to be (HLM14)
\bea
\chi_{\rm gra}(0)\approx 7.2 \times 10^{-10}\hat{f}_{p}\hat{n}_{23}\hat{T}_{\d}^{-1}~~~~,\label{eq:chi0_carb}
\ena
where $\hat{f}_{p}=f_{p}/0.1$ with $f_{p}$ being the fraction of H atoms in this case. If $f_{p}(\H)$ is too small ($\ll 0.1$), the magnetization becomes dominated by the nuclei of $^{13}C$ that has $f_{p}(^{13}C)\approx 0.01$ (see also \citealt{1999ApJ...520L..67L}). It can be seen that with $\chi_{\rm gra}(0)$, the magnetic moment due to Barnett effect for carbonaceous grains  is 
\bea
\mu_{\rm Lar,Bar}=-\frac{\chi_{\rm gra}(0)\hbar V \omega}{g_{e}\mu_{B}}\simeq 2.3\times 10^{-13}a_{-5}^{3}\omega~ \D,
\ena
which is the same order as $\mu_{\rm rot}$. The Larmor precession timescale is estimated to be
\bea
\tau_{\rm Lar,Bar}\simeq 1.45\times 10^{4}\frac{a_{-5}^{2}}{\hat{f}_{p}\hat{n}_{23}\hat{T}_{d}\hat{B}}~\yr.
\ena

According to LH07, the precession timescale of typical aligned grains along the anisotropic radiation direction (see also \citealt{LAH15}) is
\bea
\tau_{\rm RAT}&\simeq&3.2\times 10^{3}\hat{\rho} a_{-5}\frac{1.2\mum}{\bar{\lambda}}\frac{u_{\rm ISRF}}{\gamma u_{\rad}}\frac{0.01}{Q_{e3}} \yr.~~~
\ena

{ For strong, highly anisotropic radiation field (e.g., $\gamma u_{\rad}> u_{\rm ISRF}$), we have $\tau_{\rm RAT}< \tau_{\rm Lar,Bar}<\tau_{\rm Lar, rot}$, for which the local direction of starlight anisotropy becomes the axis of alignment and carbonaceous grains can get aligned with respect to the local anisotropy direction of radiation field rather than with the magnetic field. }

Furthermore, as the precession in respect to the radiation direction is still relatively slow, the carbonaceous grains accelerated by turbulence in respect to the magnetic field
(\citealt{2002ApJ...566L.105L}; \citealt{2003ApJ...592L..33Y}; \citealt{Yan:2004ko}, \citealt{2009MNRAS.397.1093Y}; \citealt{Hoang:2012cx}) should experience more randomization arising from the electric $\sim {\bf B}\times {\bf v}$ field acting on grains as they experience fluctuations of grain charging (\citealt{2006ApJ...647..390W}). The expected cumulative effect is a significant decrease of the polarization arising from carbonaceous grains. 

We note, that contrary to carbonaceous grains, the silicate ones get their magnetic moment not through the charge rotation, but through much more powerful Barnett effect (see Eq. \ref{eq:muBar_sup}), which makes the alignment with the magnetic field less susceptible to randomization by dipole fluctuations as in \cite{2006ApJ...647..390W}. {  \cite{1986ApJ...308..281M} speculated that iron inclusions may not be incorporated into big carbon grains. If true, then the additional stabilization can be applicable only for silicate grains.}

\section{Discussion}\label{sec:disc}

This work extends and elaborates our previous works (LH07; LH08; HL08) on grain alignment for dust grains with enhanced magnetic susceptibilities. We feel that it is important to stress that this work is based on RATs, and it cannot be treated as the extension of paramagnetic alignment of grains with magnetic inclusions. Indeed, RATs produce not only suprathermal rotation but also grain alignment. The role of enhanced magnetic susceptibilities is to stabilize the high-J attractor point, which allows MRAT alignment to achieve better/perfect alignment. 

\subsection{Comparison to our previous theoretical works}
The present work extends our previous detailed quantitative study of RAT alignment of ordinary paramagnetic grains (LH07; HL08) to grains with enhanced magnetic susceptibility due to magnetic inclusions. The latter grains were shown by LH08 to get better alignment as the presence of magnetic inclusions allows grains to be aligned with high angular momentum. The present study elaborates and extends that in LH08. In particular,  while LH07 derived a parameter space ($q^{\max},\psi$) for the existence of high-J and low-J attractor points for ordinary paramagnetic material,  here we have derived the similar parameter space for high-J attractors in the presence of the  enhanced magnetic relaxation. For the practical applications of the results in the paper, it is important that we identified the critical values of the ratio of the rate of magnetic relaxation to the rate of gas damping, $\delta_{\rm m,cri}$, that produce high-J attractors. 

In this paper, we present the degree of grain alignment obtained by numerical simulations that treat MRAT alignment with random excitations by gas collisions and magnetic relaxation for a wide range of physical parameters. We obtained the realistic degrees of alignment for grains of arbitrary magnetic susceptibility, including ordinary paramagnetic and superparamagnetic material. We note that earlier works on alignment with magnetic inclusions (\citealt{1995ApJ...445L..63M};  \citealt{1995ApJ...455L.181G}; \citealt{2006A&A...448L...1D}) deal with thermally rotating grains. Interestingly enough, the alignment of grains with enhanced magnetic susceptibilities rotating due to suprathermal torques suggested by \cite{1979ApJ...231..404P} can be perfect, in analogy to RATs. However, the torques in \cite{1979ApJ...231..404P} are fixed in the grain frame and do not align grains by themselves. This is the principal difference between them and RATs. The latter act in the frame related to the magnetic field and are known to induce grain alignment. The contribution of the magnetic torques that can be significantly weaker than RATs is to tip the balance to prevent the attractor point from transferring into a repellor point as $q^{\max}$ and $\psi$ change.

\subsection{Small size cut-off for aligned grains}

Inverse modeling of interstellar polarization curves for the interstellar diffuse medium shows that good alignment stops at the grain size of approximately $0.09\mum$ and is related with the inefficiency of RATs to align grains significantly smaller than the wavelength of the radiation (see Figures 30 and 31 in LH07). The residual alignment can be explained as the result of the ordinary Davis-Greenstein alignment as quantitatively demonstrated in \cite{2014ApJ...790....6H}.

Our present study does not change the conclusions related to the small size cut-off of grain alignment. Indeed, our calculations show that the presence of strongly magnetic impurities only marginally changes the threshold value for the cut-off. The alignment of thermally rotating small grains by the Davis-Greenstein process is found to be inefficient, even for superparamagnetic inclusions (HL16). However, if the small grains have superparamagnetic inclusions, this can affect the estimates of the magnetic field strength {through UV polarization} that were suggested in HLM14. In particular, the values of the magnetic field strength in the diffuse medium may be smaller than $\sim 10\mu$G, as estimated by HLM14 for ordinary paramagnetic material.

\subsection{Degree of radiative alignment for grains with magnetic inclusions}

To study the dynamics of grains we used the machinery developed in our earlier studies (see LH07, HL08). By combining the Langevin equation and the equation of steady motion by RATs, we treated the grain dynamics both during the thermally rotating and suprathermal rotation stages. We carried out simulations of MRAT alignment and obtained the degree of alignment for a wide range of physical parameters. We quantified how the degree of RAT alignment tends to increase with increasing the magnetic susceptibility $\delta_{\rm m}$. In particular, we demonstrate that the alignment of relatively big interstellar grains (e.g., $a=0.1, 0.2\mu$m) is perfect for superparamagnetic grains of $\delta_{\rm m}\sim 10$. Above this value, the alignment becomes independent of the value of $q^{\max}$ and $\psi$. This corresponds to the situation when RATs mainly play a role of spinning up grains to the suprathermal rotation, while magnetic relaxation rapidly aligns grains. 

Smaller grains ($a=0.05, 0.08\mum$) cannot reach the perfect alignment, even when the magnetic relaxation is fast of $\delta_{\rm m}>10$. { The degree of alignment $R\sim 0.02$ computed here for the $a=0.05\mu$m grains is consistent with the results obtained from inverse modeling of observed polarization in the ISM (\citealt{1995ApJ...444..293K}; \citealt{Draine:2009p3780}; HLM14)}.

\cite{1986ApJ...308..281M} (M86) suggested that the grains smaller $0.09\mum$ are not aligned because they may not have magnetic inclusions. \cite{1995ApJ...455L.181G} pointed out that the spacing frequency of superparamagnetic inclusions obtained from spectroscopy is consistent with the M86 model. Our obtained results indicate that even with magnetic inclusions, grains smaller than $0.08\mum$ cannot be efficiently aligned because RATs are so weak for these grains that cannot drive their alignment. 

We would like to stress that our scenario is very different from that by M86. The M86 model is based on the Davis-Greenstein alignment, rather than on the RAT alignment. His idea is that the probability of a grain to have magnetic inclusions increases with the size of the grain. We showed that the cut-off is not fixed by the grain properties, but the radiation field. Therefore, contrary to the suggestion in M86 we expect the variations of the small size cut-off depending on the variations of the spectrum and intensity of the radiation field. The good correspondence of the observations to the predictions based on the RAT efficiency also means that the suprathermal spin-up torques of small grains are subdominant. The thermal flipping and thermal trapping of such grains (\citealt{1999ApJ...516L..37L}; HL09b) are likely to play a role for this. 

We also would like to note, that, unlike DW96, the model of MRATs is based on the anisotropic radiation alignment. The radiative torques not only do spin up grains but also align them. The principal role of enhanced magnetic relaxation is to stabilize the high-$J$ attractor point. {The critical value of $\delta_{\rm m,cri}$ does not depend on the strength of RATs as well as local conditions. However, since the $\delta_{\rm m}\propto 1/n_{\H}$ (Eq. \ref{eq:deltam}), to have stable alignment in dense clouds, larger susceptibilities are required, compared to the value needed for the diffuse ISM.}

\subsection{Implications for iron abundance in silicate: $\phi_{\rm sp}$ and $N_{\rm cl}$}

From Figure \ref{fig:deltam_cri}, we can see that the high-J attractor points are present for the parameter space if $\delta_{\rm m}>10$. As shown in Figure \ref{fig:delta_m} (left), it can be satisfied with $\sim 10\%$ of Fe depleted into the dust (i.e.,$\phi_{\rm sp}\sim 0.03$) for $N_{\rm cl}>1$. Lower Fe depletion of $5\%$ (i.e., $\phi_{\rm sp}\sim 0.015$ can also be achieved $N_{\rm cl}>10$. 

{ Based on observed microwave emission at $90\GHz$, DL99 suggested that no more than 5$\%$ of Fe can be depleted into dust { in the form of iron clusters}. A recent study by \cite{2013ApJ...765..159D} where a new form of magnetic susceptibility is derived shows that 100$\%$ of Fe abundance can be incorporated into the grain without violate observed optical-IR extinction curves. We note that the difference in the magnetic susceptibility between DL99 and DH13 is only important for rotation frequencies of $\nu \sim 10-300$GHz (see HL16), which does not affect our analysis in Section \ref{sec:grainmag} for large grains with rotation frequency $\nu\ll 10$ GHz.} In particular, the typical value from observations of GEMS is $\phi_{\sp}=0.03$ (\citealt{Bradley:1994p6379}; \citealt{1995ApJ...445L..63M}), which indicates about $10\%$ of iron depleted into dust silicate.

The number of iron clusters per grain is given by Equation (\ref{eq:Ncl}). To have $\mathcal{N}\ge 1$, it follows that $N_{\rm cl}< 3.5\times 10^{8}\phi_{\rm sp}a_{-5}^{3}$. Therefore, even {one iron cluster} incorporated in the grain, and with $\phi_{\rm sp}<0.03$ (i.e., $10\%$ of Fe abundance), it is sufficient to produce $\delta_{\rm m}>10$ (i.e., high-J attractor points). Yet, the question is whether iron exists in the form of iron clusters or diffusely distributed within the grain, and whether iron nanoparticles are in the form of free-fliers or inclusions still remained. {Lastly, a recent analysis of local interstellar dust grains captured by Cassini satellite \citep{Altobelli:2016dl} reveals the presence of iron inclusions in silicate grains.}

\subsection{Implication for observations in the diffuse medium}

Recent Planck results \citep{2015A&A...576A.104P} show a high polarization degree up to $20\%$ for the diffuse interstellar medium (ISM). Inverse modeling of starlight polarization usually requires perfect alignment of silicate grains to reproduce the maximum possible polarization (\citealt{Draine:2009p3780}; HLM14). This raises some challenge for the traditional RAT mechanism described in LH07a. The results there show that the RAT alignment can easily reach perfect if the high-J attractor point is present (also HL08). In contrast, without high-J attractor point, the degree of alignment is only $20-30\%$. This, by itself, does not disqualify RATs from explaining the observed alignment. First of all, the existence of the high-J attractor point depends on the $q^{\max}$ parameter of grains, but its distribution is uncertain for irregular interstellar grains. Moreover, in the absence of high-J attractor points, the degree of alignment still can be significantly increased if grains are also subject to suprathermal torques and the evidence of this effect has been claimed in observations (\citealt{2013ApJ...775...84A}). The MRAT scenario elaborated in this work provides an alternative explanation. The truth should be established through more observations. 

If MRAT scenario is confirmed, this will have important consequences for interpreting the observational data. The joint action of RATs and strongly enhanced magnetic relaxation is expected to provide perfect alignment with the short grain axis parallel to the magnetic field. Second, it gives universal alignment of high efficiency in the diffuse ISM, which is naturally consistent with the Planck polarization data for the diffuse medium. 

{ Let us outline a few predictions of MRAT that can be tested for the role of magnetic torques on grain alignment. 


First, MRAT alignment can be perfect, resulting in higher polarization fraction than in RAT alignment. Thus, quantitative measurements of dust polarization is particularly useful.

Second, large grains tend to have weaker angle dependence than smaller ones because they are perfectly aligned with the magnetic field. So, the polarization fraction at longer wavelengths is expected to have weaker angle dependence than at shorter wavelengths because large grains emit dominantly at long wavelengths.

Third, the angle dependence of MRAT alignment is reduced when increasing $u_{\rad}$ because grains are suprathermally rotating and perfectly aligned. Thus, MRAT predicts a reduction in the angle dependence of polarization near the radiation source. Similarly, in environment conditions where pinwheel torques are important, we also predict the weaker angular dependence of the polarization fraction.

}
 
\subsection{Implications for observations of dust polarization in molecular clouds}

Let us discuss now the implications of our present results for current observations of dust polarization from starless cores and reflection nebula. Recent detailed observational studies of grain alignment provided evidence in favor of the traditional RAT alignment (see \citealt{Andersson:2015bq} for a review).  Below we analyze whether the observational evidence excludes the possibility of MRAT alignment that we have studied in this paper. 

First of all, it was found that the alignment can be enhanced in the presence of H$_2$ torques (\citealt{2013ApJ...775...84A}; \citealt{2015MNRAS.448.1178H}). This does not contradict to MRAT process, as the effect of H$_{2}$ torques on grain alignment is expected to be still applicable because the small size cut-off will shift to a smaller size due to H$_{2}$ torques, allowing more grains to be aligned and thus increasing the total polarization. 

Second,  \cite{2012A&A...541A..52V} found that the alignment is not correlated with the Fe depletion, which seemed to exclude the effect of superparamagnetism and the effect of enhanced relaxation. Is this so, however? Our present study shows the saturation of alignment for $\delta_{\rm m}>\delta_{\rm m,cri}$, which indicates that the fraction of Fe in the gas phase would not be good tracer of alignment. Thus, a weak correlation of the starlight polarization with dust-phase Fe abundance \citep{2012A&A...541A..52V} does not necessarily reveal that Fe is useless for grain alignment. Instead, it indicates that to some level, the increase of Fe in the dust does not increase the alignment and then the degree of polarization saturates. The formation of iron clusters is a not far- fetched assumption since above $90\%$ of Fe is present in the dust (see Figure 4 in \cite{2012A&A...541A..52V}), which is more than sufficient for magnetic alignment saturation. 

Third, observations by \cite{2011A&A...534A..19A} and \cite{2015ApJ...812L...7V} confirmed the theoretical prediction in LH07 on the dependence of RAT alignment on the anisotropy direction of the radiation. In view of our present study, only large grains of $a=0.2\mum$ that have very strong magnetic response demonstrate non-angle dependence alignment. Grains of size $a\le 0.1\mum$ can still maintain the angle dependence alignment even at largest magnetic susceptibilities because of the dependence of the maximum rotational rate $J_{\max}$ on the anisotropy direction $\psi$. {If the alignment is observed to decrease with increasing the angle $\psi$, it would also rule out the grain shape with RATs of $q^{\max}<1$, because we predict the alignment tends to increase with $\psi$ for the case of moderate magnetic susceptibility $\delta_{\rm m}<30$.} This study provides, however, the most stringent limitations on the magnetic properties of the grains. We feel that it is necessary to continue the tests like those in \cite{2011A&A...534A..19A} in order to understand these constraints. In particular, it can narrow down the RAT parameter space.  

\subsection{Toward a physical modeling of  foreground polarization for CMB polarization missions}

The size-dependence degree of grain alignment is an essential ingredient for modeling the polarization spectrum of thermal dust emission and MDE. We have obtained such a size-dependence degree of alignment from simulations using AMO, which relates directly the degree of alignment with the magnetic susceptibility (e.g., $\phi_{\sp}$), radiation fields, and local environmental conditions. It will be used to construct a self-consistent model of magnetic dipole emission and polarization spectrum because both MDE and its polarization depend on $\phi_{\sp}$.

If MRATs are responsible for most of the interstellar alignment, this can simplify the modeling of polarized emission from interstellar dust. Indeed, this type of alignment is much more robust and does not depend on the insufficiently constrained distribution of $q^{\max}$ of interstellar grains. 

This research has important implications for CMB studies. Apart from being the dominant source of polarized thermal emission via electric dipole mechanism for the frequencies larger than $300$ GHz, it can also be responsible for polarized MDE at lower frequencies (DL99, \citealt{2013ApJ...765..159D}; \citealt{2016ApJ...821...91H}). The latter emission mechanism is closely related to magnetic inclusions in grains, which are the inclusions that are invoked for the MRAT alignment. This emission, if its magneto-dipole nature is confirmed, can be used to constrain the abundance and properties of the magnetic inclusions that are appealed to within the MRAT mechanism. 

\section{Summary}\label{sec:sum}
Magnetic inclusions in interstellar dust grains have been discussed through the long history of dust research, especially in relation to the problems of grain alignment (see 
\citealt{1951ApJ...114..187S}; \citealt{Jones:1967p2924}). They were also involved in LH08 as a possible way to enhance the efficiency of the RAT mechanism. This paper elaborates this suggestion and comes to important conclusions listed below:

\begin{itemize}

\item[1] We find that even a small inclusion of iron particles can enhance the magnetic relaxation rate to above the gaseous damping rate, i.e., $\delta_{\rm m}> 1$, for which the effect of magnetic relaxation must be accounted for in the radiative alignment process. The mechanism of alignment is {\it different} from the Davis-Greenstein alignment, as the strength of radiative torques aligning grains typically significantly exceeds the strength of the torques arising from superparamagnetic relaxation. The role of the latter is to stabilize the high-J attractor points for the phase trajectories induced by RATs. Those points result in perfect alignment of grains. Also, enhanced magnetic susceptibility results in faster Larmor precession, which favors the magnetic field as an axis of grain alignment.

\item[2] Using our analytical model (AMO) of RATs, we study the MRAT alignment of enhanced magnetic grains.  We first derive the critical values of the magnetic relaxation, $\delta_{\rm m,cri}$, to produce the high-J attractor points as a function of the radiation direction $\psi$ and $q^{\max}$. For $a=0.1\mum$, we find that the high-J attractor is universal for $\delta_{\rm m}\ge 10$ for $q^{\max}=0.1-10$. This criterion perhaps requires the inclusion of even an iron cluster as small as 20 iron atoms per cluster into the silicate grains, assuming a typical filling factor $\phi_{\rm sp}=0.03$. This result does not depend on the value of RATs.  We solve the equation of steady motion by RATs and magnetic relaxation and find that the high-J attractor points are indeed present for $\delta_{\rm m}>\delta_{\rm m,cri}$.

\item[3] We carry out numerical simulations of MRAT alignment taking into account stochastic collisional excitations and magnetic fluctuations for arbitrary values of $\delta_{\rm m}$. Similar to earlier studies in HL08, we observe the transfer of grains from the low-J attractor point to the high-J attractor point as a result of collisional excitations. We numerically demonstrate that the alignment with high-J attractor points is very efficient/perfect.

\item[4] We compute the degree of MRAT alignment for a number of radiation angles $\psi$, grain sizes, and a broad range of $\delta_{\rm m}$, spanning ordinary to superparamagnetic cases. We demonstrate the dependence of alignment degree on $\psi$. We find that the degree of alignment increases with increasing the grain size $a$. In the case of superparamagnetism of $\delta_{\rm m}\ge 30$, the MRAT alignment becomes perfect for large grains of $a> 0.1\mu$m, which is independent of $\psi$ and $q^{\max}$. The size-dependence degree of alignment paves the way for physical modeling the polarization of thermal dust emission as well as magnetic dipole emission. 

\item[5] Our work calls for more tests of grain alignment in order to find out if the enhancement of magnetic relaxation is required for RATs to be able to explain the observed polarization. This is important for quantitative studies of magnetic fields in various astrophysical environments, e.g. the interstellar medium, molecular clouds and accretion disks. This is also essential for accurate modeling of the contribution to galactic polarized foreground arising from electro-dipole and magneto-dipole emission from interstellar dust. If enhanced degree of alignment is required to reproduce Planck data, then, our present results suggest that grains should have magnetic inclusions. 

\item[6] { We find that millimeter-sized grains in the accretion disks may be aligned with the magnetic field if they are incorporated with iron clusters of $10^{5}$ iron atoms per cluster. If confirmed, this opens good prospects for
tracing magnetic fields in protoplanetary disks.}

\end{itemize}

\acknowledgments
{We thank anonymous referee for his/her insightful and detailed comments that helped us improve the clarity of our paper. T.H. acknowledges the support from the Natural Sciences and Engineering Research Council of Canada (NSERC). AL acknowledges NSF grant AST 1109295, NASA grant NNH 08ZDA0090, and support of the NSF Center for Magnetic Self-Organization.}

\appendix

\section{A. AMO and RATs components}\label{apdx:AMO}

\begin{figure*}
\includegraphics[width=0.45\textwidth]{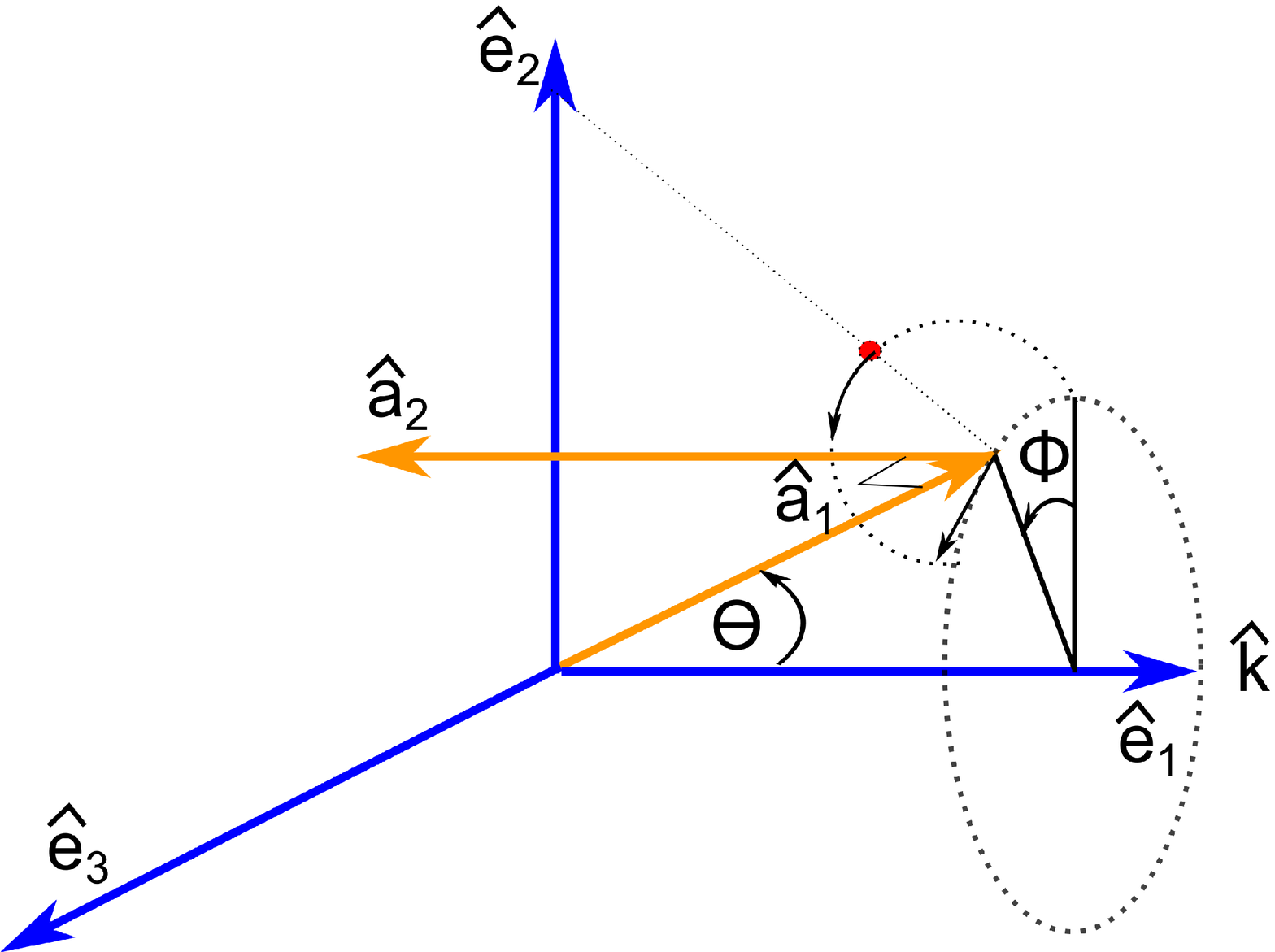}
\includegraphics[width=0.45\textwidth]{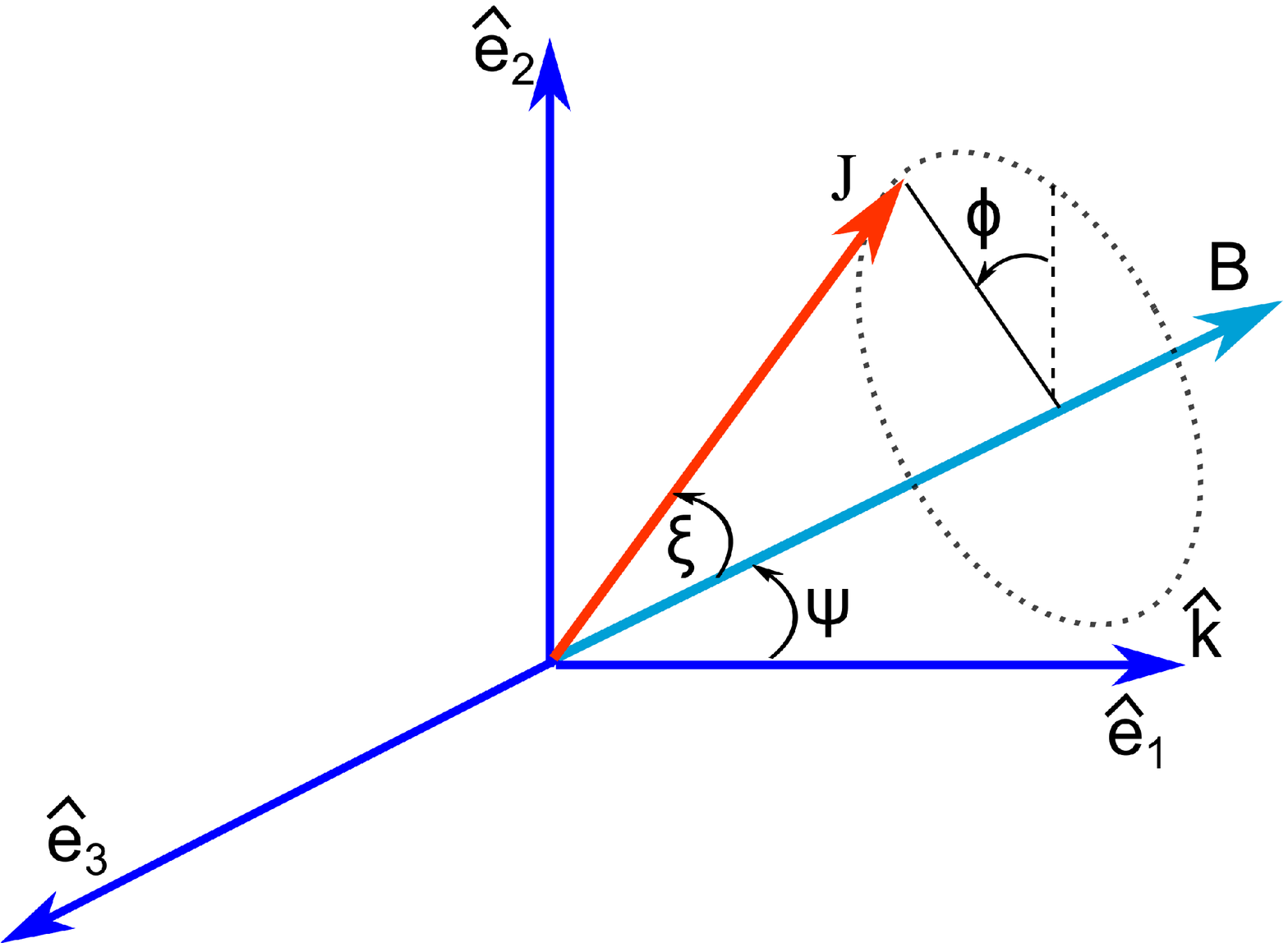}
\caption{Coordinate systems used in numerical calculations.}
\label{fig:RFs}
\end{figure*}

\subsection{A1. RATs and mean values over the spectrum of radiation fields}
The radiative torque resulting from the interaction of radiation field with a grain of size $a$ is defined by
\bea
{\bf \Gamma}_{\rad}=\frac{\gamma u_{\rad}\bar{\lambda} a^{2}}{2}\overline{\bf Q}_{\Gamma},\label{rat0}
\ena
where $\gamma$, $\bar{\lambda}$, and $u_{\rad}$ are the degree of anisotropy, mean wavelength and energy density of radiation field, respectively. Here  ${\bf Q}_{\Gamma}$ is the radiative torque efficiency vector, and overlines denote the averaging over the spectrum of the incident radiation field. 

Let $u_{\lambda}$ be the energy density per unit wavelength $\lambda$, then $\bar{\lambda}$ and $u_{\rad}$ are defined as
\bea
\overline{\lambda}=\frac{\int u_{\lambda} \lambda d\lambda}{\int u_{\lambda} d\lambda},~{u}_{\rad}=\int u_{\lambda} d\lambda. \label{urad}
\ena

Let $Q_{\Gamma}$ be the radiative torque efficiency produced by the monochromatic radiation field with wavelength $\lambda$, the mean torque efficiency over the radiation field is 
\bea
\overline{Q_{\Gamma}}=\frac{\int Q_{\Gamma}u_{\lambda} d\lambda}{\int u_{\lambda} d\lambda}\label{torqmean}.
\ena

\subsection{A2. AMO: General expressions of RATs}
{ LH07 proposed an analytical model (AMO) of RATs. This model is based on a helical grain interacting with the anisotropic radiation beam. The helical model consists of a spheroidal body and a masseless mirror attached to the body (see LH07 for details).}

RAT from the AMO is given by the similar form as Equation (\ref{rat0}):
\bea
{\bf \Gamma}_{\rad}&=&\frac{\gamma u_{\rad}\bar{\lambda} l_{2}^{2}}{2}\overline{\bf Q}_{\Gamma},\label{a1}
\ena
where $l_{2}$, defined as the grain size $a$, is the size of the squared mirror.

Using the self-similar scaling of the magnitude of RATs obtained for irregular grains of size $a$ induced by the radiation field of wavelength $\lambda$,
\bea
\left|Q_{\Gamma}\right|&\sim& 0.4\left(\frac{\lambda}{a}\right)^{-3} \mbox{ for~$\lambda > 2a$},\label{a2}\\
&\sim& 0.4 \mbox { for~$\lambda \le 2 a$},\label{a3}
\ena
and the functional forms of RATs from the AMO, we can write RAT components as follows:
\bea
Q_{e_{1}}(\Theta, \beta,\Phi= 0)&=&\frac{\left|Q_{\Gamma}\right|q^{\max}}{\sqrt{(q^{\max})^{2}+1}}\frac{q_{e_{1}}(\Theta, \beta, \Phi=0)}{q_{e_{1}}^{\max}},\label{a4}\\
Q_{e_{2}}(\Theta, \beta, \Phi=0)&=&\frac{\left|Q_{\Gamma}\right|}{\sqrt{(q^{\max})^{2}+1}}\frac{q_{e2}(\Theta, \beta, \Phi=0)}{q_{e_{2}}^{\max}},\label{a5}\\
Q_{e_{3}}(\Theta, \beta, \Phi=0)&=&\frac{\left|Q_{\Gamma}\right|q^{\max}}{\sqrt{(q^{\max})^{2}+1}}\frac{q_{e3}(\Theta, \beta, \Phi=0)}{q_{e_{3}}^{\max}},\label{a6}
\ena
where
\bea
q_{e_{1}}(\Theta, \beta, \Phi=0)&=&-\frac{4l_{1}}{\lambda}C\left(n_{1}n_{2}\frac{[3\cos^{2}\Theta-1]}{2}+\frac{n_{1}^{2}}{2}\cos\beta\sin^{2}\Theta 
-\frac{n_{2}^{2}}{2}\cos\beta\sin^{2}\Theta-\frac{n_{1}n_{2}}{2}\cos^{2}\beta\right),\label{a7}\\
q_{e_{2}}(\Theta, \beta,\Phi= 0)&=&\frac{4l_{1}}{\lambda}C\left(n_{1}^{2}\cos\beta\cos^{2}\Theta-\frac{n_{1}n_{2}}{2}\cos^{2}\beta\sin^{2}\Theta-\frac{n_{1}n_{2}}{2}\sin^{2}\Theta +n_{2}^{2}\cos\beta\sin^{2}\Theta\right),\label{a8}\\
q_{e_{3}}(\Theta, \beta,\Phi= 0)&=&\frac{4l_{1}}{\lambda}C n_{1}\sin\beta \left[n_{1}\cos\Theta-n_{2}\cos\beta\sin\Theta\right]+\left(\frac{b}{l_{2}}\right)^{2}\frac{2e a}{\lambda}(s^{2}-1)K(\Theta)\sin 2\Theta,\label{a9}
\ena
with $q_{e_{j}}^{\max}=max\langle q_{e_{j}}(\Theta, \beta,\Phi= 0)\rangle_{\beta}$ for $j=1,2$ and $3$. 
The magnitude ratio of torque components $q^{\max}$ is defined by
\bea
q^{\max}=\frac{max{\langle Q_{e_{1}}(\Theta, \beta, \Phi=0)\rangle_{\beta}}}{max{\langle Q_{e_{2}}(\Theta, \beta, \Phi=0)\rangle_{\beta}}}.\label{qmax}
\ena

In Equations (\ref{a7})-(\ref{a9}), $C$ is a function given by
\bea
C=\left|n_{1}\cos\Theta-n_{2}\sin\Theta\cos\beta\right|,
\ena
where $\Theta$ is the angle between the axis of major inertia ${\bf a}_{1}$ and the radiation direction ${\bf k}$, $\beta$ is the angle describing the rotation of the grain about $\ba_{1}$; $n_{1}=-\sin i, n_{2}=\cos i$ are components of the normal vector
of the mirror tilted by an angle $i$ in the grain coordinate system. The second term of Equation (\ref{a9}) represents the torque due to the spheroid, which is shown to result in the grain precession only (see LH07 for details).

We adopt the AMO with $i=45^{\circ}$ in this paper, unless mentioned otherwise. We also assume the amplitude of $Q_{e_{3}}$ is comparable to that of $Q_{e_{1}}$ and $Q_{e_{2}}$.

RATs at a precession angle $\Phi$ can be derived from RATs at $\Phi=0$ using the coordinate system transformation, as follows:
\bea
Q_{e_{1}}(\Theta, \beta, \Phi)&=&Q_{e_{1}}(\Theta, \beta, \Phi=0),\label{aeq4}\\ 
Q_{e_{2}}(\Theta, \beta, \Phi)&=&Q_{e_{2}}(\Theta, \beta,\Phi=
0)\mbox{cos}\Phi+Q_{e_{3}}(\Theta, \beta,\Phi=
0)\mbox{sin}\Phi,\label{aeq5} 
\\  
Q_{e_{3}}(\Theta, \beta, \Phi)&=&Q_{e2}(\Theta, \beta,\Phi=
0)\mbox{sin}\Phi-Q_{e_{3}}(\Theta, \beta,\Phi=
0)\mbox{cos}\Phi.\label{aeq6}
\ena

To study the alignment of the angular momentum with respect to magnetic field, we use the  spherical coordinate $J,\xi$ and $\phi$ (see Fig. \ref{fig:RFs}). In this coordinate system, RATs components are given by
\bea
F(\psi,\phi,\xi)&=&Q_{e_{1}}(\xi, \psi, \phi)(-\mbox{sin }\psi \mbox{cos }\xi \mbox{cos }\phi-\mbox{sin  }\xi \mbox{cos }\psi)+Q_{e_{2}}(\xi, \psi, \phi)(\mbox{cos }\psi \mbox{cos }\xi \mbox{cos }\phi-\mbox{sin }\xi
\mbox{sin }\psi)\nonumber\\
&&+Q_{e_{3}}(\xi, \psi, \phi)\mbox{cos }\xi \mbox{sin }\phi,\label{eeq9}\\ 
G(\psi,\phi,\xi)&=&Q_{e_{1}}(\xi, \psi, \phi)\mbox{sin }\psi \mbox{sin }\phi-Q_{e_{2}}(\xi, \psi, \phi)\mbox{cos }\psi \mbox{sin }\phi+Q_{e_{3}}(\xi, \psi, \phi)\mbox{cos }\phi,\label{eeq10}\\ 
H(\psi,\phi,\xi)&=&Q_{e_{1}}(\xi, \psi, \phi)(\mbox{cos }\psi \mbox{cos }\xi -\mbox{sin }\psi \mbox{sin  }\xi \mbox{cos }\phi)+Q_{e_{2}}(\xi, \psi, \phi)(\mbox{sin}\psi \mbox{cos }\xi + \mbox{cos }\psi\mbox{sin }\xi\mbox{cos  }\phi)\nonumber\\
&&+Q_{e_{3}}(\xi, \psi, \phi)\mbox{sin }\xi \mbox{sin }\phi, \label{eeq11} 
\ena
where $Q_{e_{1}}(\xi, \psi, \phi), Q_{e_{2}}(\xi, \psi, \phi), Q_{e_{3}}(\xi, \psi, \phi)$, as functions of $\xi, \psi$ and $\phi$, are components of the RAT efficiency vector in the
lab coordinate system (see DW97; LH07a). To obtain $Q_{e_{1}}(\xi, \psi, \phi),
Q_{e_{2}}(\xi, \psi, \phi)$ and $Q_{e_{3}}(\xi, \psi, \phi)$ from ${\bf Q}_{\Gamma}(\Theta, \beta, \Phi)$, we need to use the relations between $\xi, \psi, \phi$ and $\Theta, \beta, \Phi$ (see \citealt{2003ApJ...589..289W}; HL08).

\section{B. Rotational Excitation by Random Gas Collisions} \label{apdx:collexc}
For numerical estimates, we adopt the oblate spheroid grains with semi-minor and semi-major axes of $a_{1}$ and $a_{2}$, which has axial ratio $s=a_{1}/a_{2}$. The principal moments of inertia { for the rotation parallel and perpendicular to the grain symmetry axis} take the following forms:
\bea
I_{\|}=\frac{8\pi}{15}\rho a_{1}a_{2}^{4},~I_{\perp}=\frac{4\pi}{15}\rho a_{2}^{2}a_{1}\left(a_{1}^{2}+a_{2}^{2}\right).\label{eq:Iparperp}
\ena

Similar to R93, diffusion coefficients are first calculated in the grain body system, and then transformed to the lab system. The diffusion coefficients are averaged over the precession of $\hat{\ba}_{1}$ around ${\bf J}$ and over the Larmor precession angle of ${\bf J}$ about ${\bf B}$, given by 
\bea
\langle{\Delta J\rangle}_{i}&=&-\frac{J_{i}}{\tau_{\gas}} {~\rm for~ i=x,y,z},\label{eq27}
\ena
where $t_{\gas}$ is the gaseous damping time. For a spheroidal grain, the gaseous damping time is 
\bea
\tau_{\gas}&=&\frac{3}{4\sqrt{\pi}}\frac{I_{\|}}{n_{\H}mv_{th}a_{2}^{4}\Gamma_{\parallel}(e)},\label{eq28b} 
\ena
where $v_{\rm th}=\sqrt{2kT_{\gas}/m}$ is the thermal velocity of atom, $\Gamma_{\parallel}, \Gamma_{\perp}$ are factors characterizing the geometry of grain given by 
\bea
\Gamma_{\parallel}(e) &=& \frac{3}{16}{3+4(1-e^{2})g(e)-e^{-2}[1-(1-e^{2}\
)^{2}g(e)]},\label{eq29b}\\ 
\Gamma_{\perp}(e) &=& \frac{3}{32}[[7-e^{2}+(1-e^{2})g(e)+(1-2e^{-2}){(1+e^{-2})^{2}[1-(1-e^{2})g(e)]}]].\label{eq30b} 
\ena
$g(e)$ is related to the eccentricity of the grain through the expression:
\bea
g(e)=\frac{1}{2e}\mbox{ln}\left(\frac{1+e}{1-e}\right),\label{eq31} 
\ena
where $e=\sqrt{1-(a_{1}/a_{2})^{2}}$. {The limiting values are $\Gamma_{\|}=\Gamma_{\perp}=1$ for $e=0$ (i.e., spherical grains), and $\Gamma_{\|}=\Gamma_{\perp}=3/8$ for $e=1$.}

Diffusion coefficients are diagonal and given by the following expressions in the lab coordinate system in which the $z$ axis is along the magnetic field (R93) 
\bea
\langle(\Delta J_{x})^{2}\rangle&=&\frac{\sqrt{\pi}}{3}n_{\H}m^{2}a_{2}^{4}v_{th}^{3}(1+\frac{T_{d}}{T_{g}})
[(1+\cos^{2}\xi)\Gamma_{\perp}+\mbox{sin}^{2}\xi\Gamma_{\parallel}],\label{eq32}\\ 
\langle(\Delta J_{y})^{2}\rangle&=&\langle(\Delta
J_{x})^{2}\rangle,\label{eq33}\\ 
\langle(\Delta 
J_{z})^{2}\rangle&=&\frac{2\sqrt{\pi}}{3}n_{\H}m^{2}a_{2}^{4}v_{th}^{3}(1+\
\frac{T_{d}}{T_{g}})[\sin^{2}\xi\Gamma_{\perp}+\cos^{2}\xi\Gamma_{\parallel}].\label{eq34}
\ena 
Note that the above diffusion coefficients are derived by assuming perfect internal alignment of ${\bf a}_{1}$ with ${\bf  J}$ and for spheroidal grains. However, for the sake of simplicity, we can adopt these diffusion coefficients for studying the influence of gas bombardment on the alignment of irregular grains.

{ 
The excitation coefficients in the dimensionless units of $J'\equiv J/I_{\|}\omega_{T}$ and $t'\equiv t/\tau_{\H,\|}$ due to gas collisions are then given by
\bea
B'_{{\rm coll},i}=B_{{\rm coll},i}\times \left(\frac{\tau_{\gas}}{2I_{\|}kT_{\gas}} \right) {\rm for~ i=x,y,z},
\ena
where $B_{\rm coll,i}$ are given by Equations (\ref{eq32})-(\ref{eq34}).

Similarly, for magnetic fluctuations, we have the components $B'_{{\rm mag},xx}=B'_{{\rm mag},yy}=\delta_{\rm m}T_{d}/T_{\gas}$, and $B'_{{\rm mag},zz}=0$. The total excitation is then  are ${\bf B'}={\bf B'}_{\coll} + {\bf B'}_{\rm mag}$ where ${\bf B}$ denote the diagonal matrix with respective components.}

{ It is usually more convenient to solve the LEs in in the dimensionless units of $J'$ and $t'$. Therefore, Equation (\ref{eq:LE}) becomes 
\bea
dJ'_{i}=A'_{i}dt'+\sqrt{B'_{ii}}dw'_{i} \mbox{~for~} i= x,~y,~z,\label{eq:dJp_dt}
\ena
where $\langle dw_{i}^{'2}\rangle=dt'$ and
\bea
A'_{i}=-{J'_{i}}\left[\frac{1}{\tau'_{\gas,{\eff}}} +\delta_{m}(1-\delta_{iz})\right],~B'_{ii}= B'_{\rm coll,i}+\frac{T_{\d}}{T_{\gas}}\delta_{\rm m}(1-\delta_{iz}).\label{eq:Bii}
\ena
Above, $\delta_{iz}=1$ for $i=z$ and $\delta_{iz}=0$ for $i= x,y$, and $1/\tau'_{\gas,{\eff}}\approx 1+F_{\rm IR}$ where $F_{\rm IR}$ is the damping coefficient due to IR emission (see HL16).} 

\bibliography{ms.bbl}

\begin{thebibliography}{81}
\expandafter\ifx\csname natexlab\endcsname\relax\def\natexlab#1{#1}\fi

\bibitem[{Altobelli {et~al.}(2016)Altobelli, Postberg, Fiege, Trieloff, Kimura,
  Sterken, Hsu, Hillier, Khawaja, Moragas-Klostermeyer, Blum, Burton, Srama,
  Kempf, \& Gruen}]{Altobelli:2016dl}
Altobelli, N., Postberg, F., Fiege, K., {et~al.} 2016, Science, 352, 312

\bibitem[{Alves {et~al.}(2014)Alves, Frau, Girart, Franco, Santos, \&
  Wiesemeyer}]{2014A&A...569L...1A}
Alves, F.~O., Frau, P., Girart, J.~M., {et~al.} 2014, A\&A, 569, L1

\bibitem[{Andersson {et~al.}(2015)Andersson, Lazarian, \&
  Vaillancourt}]{Andersson:2015bq}
Andersson, B.-G., Lazarian, A., \& Vaillancourt, J.~E. 2015, Annual Review of
  Astronomy and Astrophysics, 53, 501

\bibitem[{Andersson {et~al.}(2011)Andersson, Pintado, Potter, Strai{\v z}ys, \&
  Charcos-Llorens}]{2011A&A...534A..19A}
Andersson, B.-G., Pintado, O., Potter, S.~B., Strai{\v z}ys, V., \&
  Charcos-Llorens, M. 2011, A\&A, 534, 19

\bibitem[{Andersson \& Potter(2007)}]{2007ApJ...665..369A}
Andersson, B.-G., \& Potter, S.~B. 2007, \apj, 665, 369

\bibitem[{Andersson \& Potter(2010)}]{2010ApJ...720.1045A}
Andersson, B.-G., \& Potter, S.~B. 2010, \apj, 720, 1045

\bibitem[{Andersson {et~al.}(2013)Andersson, Piirola, De~Buizer, Clemens,
  Uomoto, Charcos-Llorens, Geballe, Lazarian, Hoang, \&
  Vornanen}]{2013ApJ...775...84A}
Andersson, B.-G., Piirola, V., De~Buizer, J., {et~al.} 2013, \apj, 775, 84

\bibitem[{Barnett(1915)}]{Barnett:1915p6353}
Barnett, S.~J. 1915, Physical Review, 6, 239

\bibitem[{Bethell {et~al.}(2007)Bethell, Chepurnov, Lazarian, \&
  Kim}]{2007ApJ...663.1055B}
Bethell, T.~J., Chepurnov, A., Lazarian, A., \& Kim, J. 2007, \apj, 663, 1055

\bibitem[{Bradley(1994)}]{Bradley:1994p6379}
Bradley, J.~P. 1994, Science, 265, 925

\bibitem[{Chandrasekhar \& Fermi(1953)}]{1953ApJ...118..113C}
Chandrasekhar, S., \& Fermi, E. 1953, \apj, 118, 113

\bibitem[{Chiang \& Goldreich(1997)}]{1997ApJ...490..368C}
Chiang, E.~I., \& Goldreich, P. 1997, \apj, 490, 368

\bibitem[{Chiar {et~al.}(2006)Chiar, Adamson, Whittet, Chrysostomou, Hough,
  Kerr, Mason, Roche, \& Wright}]{2006ApJ...651..268C}
Chiar, J.~E., Adamson, A.~J., Whittet, D. C.~B., {et~al.} 2006, \apj, 651, 268

\bibitem[{Cho \& Lazarian(2005)}]{2005ApJ...631..361C}
Cho, J., \& Lazarian, A. 2005, \apj, 631, 361

\bibitem[{Cho \& Lazarian(2007)}]{2007ApJ...669.1085C}
Cho, J., \& Lazarian, A. 2007, \apj, 669, 1085

\bibitem[{Davis \& Greenstein(1951)}]{1951ApJ...114..206D}
Davis, L.~J., \& Greenstein, J.~L. 1951, \apj, 114, 206

\bibitem[{Davoisne {et~al.}(2006)Davoisne, Djouadi, Leroux, D'Hendecourt,
  Jones, \& Deboffle}]{2006A&A...448L...1D}
Davoisne, C., Djouadi, Z., Leroux, H., {et~al.} 2006, A\&A, 448, L1

\bibitem[{Dolginov \& Mitrofanov(1976)}]{1976Ap&SS..43..291D}
Dolginov, A.~Z., \& Mitrofanov, I.~G. 1976, Ap\&SS, 43, 291

\bibitem[{{Draine}(1996)}]{Draine:1996p6977}
{Draine}, B.~T. 1996, in Astronomical Society of the Pacific Conference Series,
  Vol.~97, Polarimetry of the Interstellar Medium, ed. W.~G. {Roberge} \&
  D.~C.~B. {Whittet}, 16

\bibitem[{Draine \& Fraisse(2009)}]{Draine:2009p3780}
Draine, B.~T., \& Fraisse, A.~A. 2009, \apj, 696, 1

\bibitem[{Draine \& Hensley(2013)}]{2013ApJ...765..159D}
Draine, B.~T., \& Hensley, B. 2013, \apj, 765, 159

\bibitem[{Draine \& Lazarian(1998)}]{1998ApJ...508..157D}
Draine, B.~T., \& Lazarian, A. 1998, \apj, 508, 157

\bibitem[{Draine \& Lazarian(1999)}]{1999ApJ...512..740D}
Draine, B.~T., \& Lazarian, A. 1999, \apj, 512, 740

\bibitem[{Draine \& Weingartner(1996)}]{1996ApJ...470..551D}
Draine, B.~T., \& Weingartner, J.~C. 1996, \apj, 470, 551

\bibitem[{Draine \& Weingartner(1997)}]{1997ApJ...480..633D}
Draine, B.~T., \& Weingartner, J.~C. 1997, \apj, 480, 633

\bibitem[{Duley(1978)}]{1978ApJ...219L.129D}
Duley, W.~W. 1978, \apj, 219, L129

\bibitem[{Dunkley {et~al.}(2009)Dunkley, Amblard, Baccigalupi, Betoule, Chuss,
  Cooray, Delabrouille, Dickinson, Dobler, Dotson, Eriksen, Finkbeiner, Fixsen,
  Fosalba, Fraisse, Hirata, Kogut, Kristiansen, Lawrence, Magalha~Es,
  Miville-Desch{\^e}nes, Meyer, Miller, Naess, Page, Peiris, Phillips,
  Pierpaoli, Rocha, Vaillancourt, \& Verde}]{2009AIPC.1141..222D}
Dunkley, J., Amblard, A., Baccigalupi, C., {et~al.} 2009, in CMB POLARIZATION
  WORKSHOP: THEORY AND FOREGROUNDS: CMBPol Mission Concept Study. AIP
  Conference Proceedings, 222--264

\bibitem[{Falceta-Gon{\c c}alves {et~al.}(2008)Falceta-Gon{\c c}alves,
  Lazarian, \& Kowal}]{2008ApJ...679..537F}
Falceta-Gon{\c c}alves, D., Lazarian, A., \& Kowal, G. 2008, \apj, 679, 537

\bibitem[{Girart {et~al.}(2006)Girart, Rao, \& Marrone}]{2006Sci...313..812G}
Girart, J.~M., Rao, R., \& Marrone, D.~P. 2006, Science, 313, 812

\bibitem[{Goodman \& Whittet(1995)}]{1995ApJ...455L.181G}
Goodman, A.~A., \& Whittet, D. C.~B. 1995, \apjl, 455, L181

\bibitem[{Greenberg(1968)}]{Greenberg:1968p6020}
Greenberg, J.~M. 1968, Nebulae and interstellar matter. Edited by Barbara M.
  Middlehurst; Lawrence H. Aller. Library of Congress Catalog Card Number
  66-13879. Published by the University of Chicago Press, 221

\bibitem[{Hall(1949)}]{Hall:1949p5890}
Hall, J.~S. 1949, Science, 109, 166

\bibitem[{Hildebrand(1988)}]{Hildebrand:1988p2566}
Hildebrand, R.~H. 1988, Royal Astronomical Society, 29, 327

\bibitem[{Hiltner(1949)}]{Hiltner:1949p5851}
Hiltner, W.~A. 1949, Nature, 163, 283

\bibitem[{Hoang {et~al.}(2010)Hoang, Draine, \& Lazarian}]{Hoang:2010jy}
Hoang, T., Draine, B.~T., \& Lazarian, A. 2010, \apj, 715, 1462

\bibitem[{Hoang \& Lazarian(2008)}]{Hoang:2008gb}
Hoang, T., \& Lazarian, A. 2008, \mnras, 388, 117

\bibitem[{Hoang \& Lazarian(2009{\natexlab{a}})}]{2009ApJ...697.1316H}
Hoang, T., \& Lazarian, A. 2009{\natexlab{a}}, \apj, 697, 1316

\bibitem[{Hoang \& Lazarian(2009{\natexlab{b}})}]{2009ApJ...695.1457H}
Hoang, T., \& Lazarian, A. 2009{\natexlab{b}}, \apj, 695, 1457

\bibitem[{Hoang \& Lazarian(2014)}]{2014MNRAS.438..680H}
Hoang, T., \& Lazarian, A. 2014, \mnras, 438, 680

\bibitem[{Hoang \& Lazarian(2016)}]{2016ApJ...821...91H}
Hoang, T., \& Lazarian, A. 2016, \apj, 821, 91

\bibitem[{Hoang {et~al.}(2015)Hoang, Lazarian, \&
  Andersson}]{2015MNRAS.448.1178H}
Hoang, T., Lazarian, A., \& Andersson, B.-G. 2015, \mnras, 448, 1178

\bibitem[{Hoang {et~al.}(2014)Hoang, Lazarian, \& Martin}]{2014ApJ...790....6H}
Hoang, T., Lazarian, A., \& Martin, P.~G. 2014, \apj, 790, 6

\bibitem[{Hoang {et~al.}(2012)Hoang, Lazarian, \& Schlickeiser}]{Hoang:2012cx}
Hoang, T., Lazarian, A., \& Schlickeiser, R. 2012, \apj, 747, 54

\bibitem[{Hughes {et~al.}(2009)Hughes, Wilner, Cho, Marrone, Lazarian, Andrews,
  \& Rao}]{2009ApJ...704.1204H}
Hughes, A.~M., Wilner, D.~J., Cho, J., {et~al.} 2009, \apj, 704, 1204

\bibitem[{Jenkins(2009)}]{2009ApJ...700.1299J}
Jenkins, E.~B. 2009, The Astrophysical Journal, 700, 1299

\bibitem[{Jones \& Spitzer(1967)}]{Jones:1967p2924}
Jones, R.~V., \& Spitzer, L. 1967, \apj, 147, 943

\bibitem[{Jones {et~al.}(2014)Jones, Bagley, Krejny, Andersson, \&
  Bastien}]{Jones:2014fk}
Jones, T.~J., Bagley, M., Krejny, M., Andersson, B.-G., \& Bastien, P. 2014,
  \aj, 149, 1

\bibitem[{Kim \& Martin(1995)}]{1995ApJ...444..293K}
Kim, S.-H., \& Martin, P.~G. 1995, \apj, 444, 293

\bibitem[{Lazarian(1997)}]{1997MNRAS.288..609L}
Lazarian, A. 1997, \mnras, 288, 609

\bibitem[{Lazarian(2007)}]{2007JQSRT.106..225L}
Lazarian, A. 2007, J. Quant. Spectrosc. Rad. Trans., 106, 225

\bibitem[{{Lazarian} {et~al.}(2015){Lazarian}, {Andersson}, \& {Hoang}}]{LAH15}
{Lazarian}, A., {Andersson}, B.-G., \& {Hoang}, T. 2015, in Polarimetry of
  stars and planetary systems, ed. L.~{Kolokolova}, J.~{Hough}, \& A.-C.
  {Levasseur-Regourd} (New York:Cambridge University Press), 81

\bibitem[{Lazarian \& Draine(1999{\natexlab{a}})}]{1999ApJ...520L..67L}
Lazarian, A., \& Draine, B.~T. 1999{\natexlab{a}}, \apj, 520, L67

\bibitem[{Lazarian \& Draine(1999{\natexlab{b}})}]{1999ApJ...516L..37L}
Lazarian, A., \& Draine, B.~T. 1999{\natexlab{b}}, \apj, 516, L37

\bibitem[{Lazarian \& Hoang(2007{\natexlab{a}})}]{2007MNRAS.378..910L}
Lazarian, A., \& Hoang, T. 2007{\natexlab{a}}, \mnras, 378, 910

\bibitem[{Lazarian \& Hoang(2007{\natexlab{b}})}]{Lazarian:2007p2442}
Lazarian, A., \& Hoang, T. 2007{\natexlab{b}}, \apj, 669, L77

\bibitem[{Lazarian \& Hoang(2008)}]{Lazarian:2008fw}
Lazarian, A., \& Hoang, T. 2008, \apj, 676, L25

\bibitem[{Lazarian \& Roberge(1997)}]{1997ApJ...484..230L}
Lazarian, A., \& Roberge, W.~G. 1997, \apj, 484, 230

\bibitem[{Lazarian \& Yan(2002)}]{2002ApJ...566L.105L}
Lazarian, A., \& Yan, H. 2002, \apj, 566, L105

\bibitem[{Martin(1971)}]{1971MNRAS.153..279M}
Martin, P.~G. 1971, \mnras, 153, 279

\bibitem[{Martin(1995)}]{1995ApJ...445L..63M}
Martin, P.~G. 1995, \apj, 445, L63

\bibitem[{Mathis(1986)}]{1986ApJ...308..281M}
Mathis, J.~S. 1986, \apj, 308, 281

\bibitem[{Mathis {et~al.}(1983)Mathis, Mezger, \&
  Panagia}]{1983A&A...128..212M}
Mathis, J.~S., Mezger, P.~G., \& Panagia, N. 1983, A\&A, 128, 212

\bibitem[{Matsumura {et~al.}(2011)Matsumura, Kameura, Kawabata, Akitaya,
  Isogai, \& Seki}]{2011PASJ...63L..43M}
Matsumura, M., Kameura, Y., Kawabata, K.~S., {et~al.} 2011, PASJ, 63, L43

\bibitem[{Morrish(2001)}]{Morrish:2001vp}
Morrish, A.~H. 2001, {The Physical Principles of Magnetism} (Wiley-IEEE Press)

\bibitem[{{Planck Collaboration} {et~al.}(2014){Planck Collaboration}, Adam,
  Ade, \& et~al.}]{2014arXiv1409.5738P}
{Planck Collaboration}, Adam, R., Ade, P. A.~R., \& et~al. 2014,
  arXiv:1409.5738, 5738

\bibitem[{{Planck Collaboration} {et~al.}(2015){Planck Collaboration}, Ade,
  Aghanim, Alina, \& et~al.}]{2015A&A...576A.104P}
{Planck Collaboration}, Ade, P. A.~R., Aghanim, N., Alina, D., \& et~al. 2015,
  Astronomy and Astrophysics, 576, A104

\bibitem[{Purcell(1979)}]{1979ApJ...231..404P}
Purcell, E.~M. 1979, \apj, 231, 404

\bibitem[{Purcell \& Spitzer(1971)}]{1971ApJ...167...31P}
Purcell, E.~M., \& Spitzer, L.~J. 1971, \apj, 167, 31

\bibitem[{Reissl {et~al.}(2016)Reissl, Wolf, \& Brauer}]{2016arXiv160405305R}
Reissl, S., Wolf, S., \& Brauer, R. 2016, arXiv:1604.05305, arXiv:1604.05305

\bibitem[{Roberge {et~al.}(1993)Roberge, Degraff, \&
  Flaherty}]{1993ApJ...418..287R}
Roberge, W.~G., Degraff, T.~A., \& Flaherty, J.~E. 1993, \apj, 418, 287

\bibitem[{Roberge \& Lazarian(1999)}]{1999MNRAS.305..615R}
Roberge, W.~G., \& Lazarian, A. 1999, \mnras, 305, 615

\bibitem[{Spitzer \& Tukey(1951)}]{1951ApJ...114..187S}
Spitzer, L.~J., \& Tukey, J.~W. 1951, \apj, 114, 187

\bibitem[{Vaillancourt \& Andersson(2015)}]{2015ApJ...812L...7V}
Vaillancourt, J.~E., \& Andersson, B.-G. 2015, \apjl, 812, L7

\bibitem[{Voshchinnikov {et~al.}(2012)Voshchinnikov, Henning, Prokopjeva, \&
  Das}]{2012A&A...541A..52V}
Voshchinnikov, N.~V., Henning, T., Prokopjeva, M.~S., \& Das, H.~K. 2012, A\&A,
  541, 52

\bibitem[{Weingartner(2006)}]{2006ApJ...647..390W}
Weingartner, J.~C. 2006, \apj, 647, 390

\bibitem[{Weingartner \& Draine(2001)}]{2001ApJ...548..296W}
Weingartner, J.~C., \& Draine, B.~T. 2001, \apj, 548, 296

\bibitem[{Weingartner \& Draine(2003)}]{2003ApJ...589..289W}
Weingartner, J.~C., \& Draine, B.~T. 2003, \apj, 589, 289

\bibitem[{Whittet {et~al.}(2008)Whittet, Hough, Lazarian, \&
  Hoang}]{2008ApJ...674..304W}
Whittet, D. C.~B., Hough, J.~H., Lazarian, A., \& Hoang, T. 2008, \apj, 674,
  304

\bibitem[{Yan \& Lazarian(2003)}]{2003ApJ...592L..33Y}
Yan, H., \& Lazarian, A. 2003, \apj, 592, L33

\bibitem[{Yan {et~al.}(2004)Yan, Lazarian, \& Draine}]{Yan:2004ko}
Yan, H., Lazarian, A., \& Draine, B.~T. 2004, \apj, 616, 895

\bibitem[{Yan \& Yan(2009)}]{2009MNRAS.397.1093Y}
Yan, H., \& Yan, H. 2009, \mnras, 397, 1093

\end{thebibliography}
\end{document}